\documentclass[pra,aps,superscriptaddress,twocolumn,a4paper]{revtex4}

\usepackage[english]{babel}
\usepackage{mathtools,amsmath,amsfonts,amssymb}
\usepackage[inline]{enumitem}
\usepackage{graphicx}
\usepackage[hidelinks]{hyperref}
\usepackage{adjustbox}
\usepackage{bm,dsfont}
\usepackage[normalem]{ulem}
\usepackage{bbm}
\usepackage{placeins}
\usepackage{csquotes}
\usepackage{braket}
\usepackage{xparse}
\usepackage{stackengine}\stackMath
\usepackage{cprotect}
\usepackage{microtype}

\NewDocumentCommand{\ketbra}{o m m}{%
  |#2\rangle
  \IfValueT{#1}{\kern-1pt_{#1}\kern-1pt}%
  \langle#3|%
}
\renewcommand{\vec}[1]{\bm{#1}}
\DeclareMathOperator{\tr}{{\rm tr}}
\newcommand{\FP}{{\rm FP}}
\newcommand{\FN}{{\rm FN}}
\newcommand{\oRMSZ}{\overline{\rm RMSZ}}
\newcommand{\oRMSE}{\overline{\rm RMSE}}

\begin{document}
\title{Characterizing a neutral atom quantum simulator accessible in cloud}

\author{Andrea Muratori}
\affiliation{Dipartimento di Fisica e Astronomia, Università di Bologna, 40127 Bologna, Italy}

\author{Francesco Minardi}
\affiliation{Dipartimento di Fisica e Astronomia, Università di Bologna, 40127 Bologna, Italy}
\affiliation{Istituto Nazionale di Ottica, CNR-INO, 50019 Sesto Fiorentino, Italy}
\affiliation{Dipartimento di Fisica e Astronomia and LENS, Università di Firenze, 50019 Sesto Fiorentino, Italy}

\author{Tommaso Calarco}
\affiliation{Dipartimento di Fisica e Astronomia, Università di Bologna, 40127 Bologna, Italy}
\affiliation{Forschungszentrum Jülich GmbH, Peter Grünberg Institute, Quantum Control (PGI-8), 52425 Jülich, Germany}
\affiliation{Institute for Theoretical Physics, University of Cologne, Zülpicher Straße 77, 50937 Cologne, Germany}

\author{Felix Motzoi}
\affiliation{Forschungszentrum Jülich GmbH, Peter Grünberg Institute, Quantum Control (PGI-8), 52425 Jülich, Germany}
\affiliation{Institute for Theoretical Physics, University of Cologne, Zülpicher Straße 77, 50937 Cologne, Germany}

\date{\today}

\begin{abstract}
    Quantum simulators have become a reality in recent years, with several platforms now remotely accessible through the cloud. These systems give researchers the opportunity to run quantum programs without requiring either a dedicated laboratory or extensive experimental expertise, and, in several cases, they have already enabled the execution of relevant protocols with remarkable results.
    However, despite the impressive progress in the performance of these machines, they still suffer from various noise sources, some of which are not always completely clear to the end user, which limits their applicability to tasks that are sufficiently robust against noise of various kinds.
    In this work, we investigate the behavior of {\em Fresnel}, a remotely accessible neutral-atom analog quantum simulator provided by Pasqal. We design and perform a series of experiments probing the machine’s response to various pulse sequences and experimental settings, and identify the main noise sources affecting its dynamics. Based on these data, we construct an effective model of the system’s noisy evolution and use numerical optimization to infer the noise strengths. The resulting model accurately reproduces the simulator’s dynamics, showing how effective noise models can be built for cloud-based quantum resources.
\end{abstract}

\maketitle

\section{Introduction}

In recent years, the development of quantum technologies has taken a major step forward, with experimental demonstrations of quantum computational advantage, the realization of programmable quantum simulators comprising hundreds of qubits, increasingly large-scale digital simulations of strongly correlated quantum systems, and substantial progress in quantum error correction and logical-qubit processing~\cite{Bernien2017,Zhong2020,Ebadi2021,Kim2023,Bluvstein2024,GoogleQAI2025,SalesRodriguez2025}.
Such progress has eventually led to quantum hardware becoming available outside specialist research laboratories, both through commercially supplied systems~\cite{Merali2011,juelich2025purchase,ionqquantumbasel2025,pasqal2026sol}, and through remotely accessible quantum-computing services~\cite{ibm2016cloud,awsbraket2026hardware,azurequantum2026providers}, thus allowing researchers the possibility to perform experiments on quantum hardware without operating a dedicated experimental infrastructure or possessing detailed expertise in the underlying experimental apparatus.

Among the various physical architectures available, neutral-atom platforms are particularly well suited to analog quantum simulation, namely the task of reproducing the dynamics of a target quantum model by appropriately engineering and controlling the many-body Hamiltonian of a quantum system used to implement the simulation~\cite{Georgescu2014,Browaeys2020,Bloch2012,Bernien2017}. In these systems, individual atoms are trapped in reconfigurable geometries using optical tweezers and coherently coupled between their electronic ground state and a highly excited Rydberg state.
The inherent interactions between Rydberg atoms naturally implement interacting spin Hamiltonians and enable the exploration of many-body phenomena in a broad range of geometries and parameter regimes. The possibility of arranging atoms in arbitrary two-dimensional (or potentially even three-dimensional) configurations, together with the availability of time-dependent controls, makes these devices versatile platforms for studying equilibrium and nonequilibrium quantum dynamics. In particular, these platforms have delivered remarkable results in the study of many-body dynamics~\cite{Bernien2017,Bluvstein2021,Dag2025}, phases of matter~\cite{Ebadi2021,Scholl2021,Semeghini2021}, and magnetic materials~\cite{Mendes2026,Qiao2025}, as well as in investigations of error-induced localization~\cite{Rava2025}, quantum state engineering~\cite{Balewski2024,Omran2019,Catalano2025}, and the optimization of classical problems~\cite{Ebadi2022,Tibaldi2023}.

The increasing availability of neutral-atom quantum processors introduces an important characterization and validation problem. Although users can specify almost freely the geometry of the atomic register and the time dependence of the available control fields, the dynamics implemented by the physical device may deviate from the ideal target evolution.
For instance, relevant imperfections may include errors in state preparation and measurement, driving fields distortions, or interaction with environment.
Their influence depends strongly on the target protocol and the measured observables. In some settings, such as in protocols relying on the Rydberg-blockade mechanism, the estimation of certain observables might be limitedly influenced by these noise sources~\cite{Bernien2017}. Protocols requiring precise coherent evolution, such as high-fidelity multiqubit gates and optimal-control sequences, can instead be especially sensitive to such imperfections~\cite{Evered2023,Mohan2023,Jandura2023}.
A deeper understanding of device nonidealities is therefore essential both for interpreting experimental observations and determining which tasks can be reliably implemented on a given processor. Furthermore, it also constitutes a prerequisite for the design of robust control protocols and for the development of device-characterization methods that can be applied directly on the target hardware.

Characterizing a device through cloud access, however, differs substantially from characterizing it within a laboratory. In the latter setting, researchers have direct access to the hardware and can perform dedicated calibration and diagnostic measurements in addition to executing quantum protocols. Such measurements provide detailed information about the physical origin, magnitude, and temporal behavior of individual noise sources~\cite{deLsleuc2018}. By contrast, a remote user can access only the information exposed through the cloud interface, such as declared device specifications and control constraints, together with the measurement outcomes obtained from the experiments they submit.
Remote characterization is therefore naturally formulated as a system-identification problem under limited observability: rather than reconstructing a complete microscopic description of the device, one seeks to infer an effective model from input-output data, which should reproduce the experimentally observed behavior of the physical system and identify the effective parameters governing its deviations from the ideal target dynamics.

In this work, we characterize \emph{Fresnel}, a neutral-atom analog quantum simulator made available through cloud access by the European company Pasqal~\cite{pasqal2026cloud}. As remote users and using only these input-output data, our objective is to construct an effective model capable of reproducing the experimental observations and estimating the strengths of the dominant error mechanisms. To this end, we perform experiments sampled at multiple evolution times, thereby exploiting the information encoded in the dynamical behavior of the system to recognize signatures of different noise sources.
Guided by the observations collected, we construct an effective open-system model that augments the ideal Rydberg Hamiltonian with a set of physically motivated error mechanisms to reproduce the observed experimental dynamics.
Because the magnitudes of the relevant noise parameters are not known \textit{a priori}, we estimate them by numerically optimizing the agreement between the model predictions and the experimental data, while constraining their values to physically reasonable ranges. Moreover, to account for the different timescales over which these parameters are expected to vary, we distinguish between slowly varying parameters, which are constrained to take common values across multiple experimental datasets, and more rapidly varying parameters, which are estimated independently for each dataset.

We show that the resulting model reproduces the relevant features of the experimental dynamics and closely agrees with the data returned by the quantum hardware. These results demonstrate that an effective model can capture a substantial part of the behavior of remotely accessible quantum hardware and highlight the diagnostic value of time-resolved experiments. In particular, they show that cloud-accessible measurements can reveal reproducible and physically interpretable signatures of the dominant device imperfections, even without direct access to the experimental apparatus.
At the same time, our analysis reveals an interplay among different effective noise mechanisms. Because of this, the inferred parameters should be interpreted primarily as an effective characterization of the device dynamics, rather than as a unique reconstruction of the underlying microscopic noise processes.
Although this work focuses on neutral-atom quantum simulators, the general methodology can be extended to the characterization of remotely accessible quantum devices based on other hardware architectures, including superconducting circuits and trapped ions. Such extensions would require adapting the dynamical model, parameterization, and measurement protocols to the controls, observables, and dominant imperfections of the target platform.

The remainder of this paper is organized as follows.
In Sec.~\ref{sec:hardware} we present Fresnel and the main characteristics of the hardware under study.
In Sec.~\ref{sec:noise} we discuss the noise sources relevant to a neutral-atom quantum simulator and describe their effects on the dynamics, as a basis for interpreting the experimental data and constructing an effective model. In particular, we focus on state-preparation and measurement errors (Sec.~\ref{sec:spam}), coherent errors (Sec.~\ref{sec:coherent}), and open-system errors (Sec.~\ref{sec:opensys}).
Section~\ref{sec:experiments} describes the experiments performed, which are then analyzed in Sec.~\ref{sec:investigation} in light of the noise effects discussed previously. This analysis leads to the effective-model ansatz for Fresnel presented in Sec.~\ref{sec:model}.
Section~\ref{sec:results} presents the results obtained by applying our effective model and the associated optimization routine to the experimental data.
Finally, Sec.~\ref{sec:conclusions} summarizes our findings with some concluding remarks.
Several computations and technical details are deferred to Appendices.

\section{Hardware specifications}\label{sec:hardware}
The hardware considered in this work is the neutral-atom quantum simulator \emph{Fresnel}, which is accessible through Pasqal's cloud platform~\cite{pasqal2026cloud}. The device is based on arrays of Rubidium atoms trapped and arranged by optical tweezers. Each atom encodes a physical qubit: the electronic ground state, $\ket{g}=\ket{5s_{1/2}}$, represents the computational basis state $\ket{0}$, while a highly excited electronic state, known as a Rydberg state and denoted by $\ket{r}=\ket{ns_{1/2}}$, represents the computational basis state $\ket{1}$. In the Fresnel device, the selected Rydberg level has principal quantum number $n_{\rm Ryd}=60$, corresponding to $\ket{r}=\ket{60s_{1/2}}$, and  the coherent coupling between $\ket{g}$ and $\ket{r}$ is implemented via a two-photon transition.

The (ideal) effective Hamiltonian governing the system is the Rydberg Hamiltonian with global controls only, namely
\begin{equation}
\begin{split}
    H
    &= \frac{\Omega(t)}{2}
    \sum_{j=1}^N
    \left(
        e^{-i\phi(t)}\ketbra[j]{g}{r}
        + e^{i\phi(t)}\ketbra[j]{r}{g}
    \right) \\
    &\quad
    - \delta(t)\sum_{j=1}^N \ketbra[j]{r}{r}
    + \sum_{j=1}^N\sum_{k<j}
    \frac{C_6}{r_{jk}^6}
    \ketbra[jk]{rr}{rr} \, ,
\label{eq:HRyd}
\end{split}
\end{equation}
where $N$ is the total number of atoms, $C_6$ is the van der Waals interaction coefficient between Rydberg states (which, for $n_{\rm Ryd}=60$, amounts to $C_6=8.66\times10^5$~rad\,$\mu$m$^6$/$\mu$s.), and $r_{jk}$ is the distance between the $j$-th and $k$-th atoms. The remaining fields are the Rabi frequency (also called amplitude) $\Omega(t)$, the detuning $\delta(t)$, and the laser phase $\phi(t)$.

Although Eq.~\eqref{eq:HRyd} may appear to allow $\Omega(t)$, $\delta(t)$, and $\phi(t)$ to be varied independently, these controls satisfy
\begin{equation}
    \delta(t) = -\frac{{\rm d}\phi(t)}{{\rm d}t} \, .
    \label{eq:detuning-phase}
\end{equation}
Consequently, the system has only two independent time-dependent controls, and only one of $\phi(t)$ and $\delta(t)$ can be specified independently. In our experiments, we use $\delta(t)$ as the time-dependent control field, while allowing for an overall piecewise-constant phase. Under this restriction, Eq.~\eqref{eq:HRyd} can be expressed in terms of the Pauli operators
\begin{equation*}
\begin{split}
    \sigma^x&=\ketbra rg+\ketbra gr\\
    \sigma^y&=i\ketbra rg-i\ketbra gr\\
    \sigma^z&=\ketbra gg-\ketbra rr
\end{split}
\end{equation*}
as
\begin{equation}
\begin{split}
    H
    &= \frac{\Omega(t)}{2}
    \sum_{j=1}^N
    \left(
        \cos(\phi)\sigma_j^x
        + \sin(\phi)\sigma_j^y
    \right) \\
    &\quad
    - \delta(t)\sum_{j=1}^N n_j^z
    + \sum_{j=1}^N\sum_{k<j}
    U_{jk}n_j^z n_k^z \, ,
\label{eq:HRyd2}
\end{split}
\end{equation}
which, for $\phi=0$, reduces to
\begin{equation}
    H
    = \frac{\Omega(t)}{2}\sum_{j=1}^N \sigma_j^x
    - \delta(t)\sum_{j=1}^N n_j^z
    + \sum_{j=1}^N\sum_{k<j} U_{jk}n_j^z n_k^z \, ,
\label{eq:HRyd3}
\end{equation}
where $n_j^z=\ketbra[j]{r}{r}$ is the projector onto the excited state of atom $j$, and $U_{jk}=C_6/r_{jk}^6$ is the interaction strength between atoms $j$ and $k$.

\begin{table}
    \centering
    \begin{tabular}{|c|c|c|c|c|}
        \hline
        parameter &min &max\\\hline
        $\Omega$ &$0$~rad/$\mu$s &$4\pi$~rad/$\mu$s\\\hline
        $\delta$ &$-20\pi$~rad/$\mu$s &$20\pi$~rad/$\mu$s\\\hline
        $T$ &$16$~ns &$6000$~ns\\\hline
        $r$ &$5$~$\mu$m &$92$~$\mu$m\\\hline
        $N$ &$1$ &$100$\\\hline
    \end{tabular}
    \caption{Overview of relevant parameters' constraints for Fresnel.}
    \label{tab:specs}
\end{table}

The Fresnel device supports registers of up to $N_{\rm max}=100$ atoms, which can be arranged in arbitrary two-dimensional geometries in the $xy$ plane. The minimum allowed interatomic distance is $r_{\rm min}=5$~$\mu$m, and the register must lie within a circular region of maximum radius $R_{\rm max}=46$~$\mu$m.
The Rabi frequency is constrained to the interval
$\Omega(t)\in[\Omega_{\rm min},\Omega_{\rm max}]=[0,4\pi]$~rad/$\mu$s with a minimum time-average amplitude of $\pi/10$~rad/$\mu$s,
whereas the detuning is restricted to
$\delta(t)\in[\delta_{\rm min},\delta_{\rm max}]=[-20\pi,20\pi]$~rad/$\mu$s,
and the laser carrying them propagates in the $y$ direction.
The total evolution time must satisfy
$T\in[T_{\rm min},\allowbreak T_{\rm max}]=[16,6000]$~ns, with controls discretized with a clock period of $4$~ns.
A brief sum up of the parameters' main constraints can be found in Tab.~\ref{tab:specs}.

At the end of the evolution, all atoms are measured simultaneously using the following procedure. First, the optical tweezers are switched back on. The tweezers trap atoms in the ground state $\ket{g}$, whereas atoms in the Rydberg state $\ket{r}$ are anti-trapped and consequently expelled from the register. A fluorescence image of the remaining atoms is then acquired. An occupied site, corresponding to a recaptured atom in $\ket{g}$, is assigned to the measurement outcome $0$, whereas an empty site is assigned to the outcome $1$. This procedure therefore yields samples from a projective measurement of the prepared state in the computational, or $z$, basis.

\section{Overview of noise sources and their effects}\label{sec:noise}
Errors affecting neutral-atom quantum hardware that we consider here can be generically classified in three categories: state-preparation and measurement errors, coherent errors, and open-system environment errors.
In this Section, we review the effects of various noise sources belonging to such categories on the system dynamics, with the aim to facilitate the interpretation of the results presented in the following Sections. We refer to Ref.~\cite{deLsleuc2018} as well for a helpful review of noise sources in Rydberg-atom experiments.

\subsection{State-preparation and measurement errors}\label{sec:spam}
We analyze in this Section the so-called {\em state-preparation and measurement} (SPAM) errors. Unlike other error sources, these do not happen during the system evolution itself, but, respectively, before and after it.
In particular, state-preparation errors account for imperfect preparation of the initial state of the register, while measurement errors consider imperfect detection of the state at the end of the quantum evolution. In the following, we review the two separately in detail.

\subsubsection{State-preparation errors}
State-preparation errors account for the failure to initialize all atoms in the electronic ground state at the beginning of the pulse sequence.
In particular, during the preparation of the register, some atoms may not be correctly optically pumped into $\ket{g}$. 
These atoms do not participate in the subsequent driven dynamics because they are no longer resonant with the transitions addressed by the laser; hence, for the purposes of the simulation, they can be treated as defects in the register.
Although the effect of state-preparation errors on the quantum dynamics can be readily accounted for by removing the affected atoms from the simulation (the precise modification depending on the protocol implemented), their effect on the final measurement is more ambiguous.
Indeed, two possible scenarios can be considered:
either
\begin{enumerate*}[label=(\roman*)]
    \item an atom is not optically pumped in $\ket{g}$, but in some other Zeeman or hyperfine state of the $\ket{5s_{1/2}}$ manifold (as is the case analyzed in Ref.~\cite{deLsleuc2018}), and during the measurement process, the atom is nevertheless recaptured in the optical tweezers and detected as being in $\ket{g}$; or
    \item an atom is prepared in some other state that is not recaptured, or even lost during the assembly of the resister, and is ultimately detected as being in $\ket{r}$ during the measurement process.
\end{enumerate*}
In both scenarios, state-preparation errors shift the measured state populations by an amount that depends on the preparation-error rate.

Let $\eta$ denote the probability that a given atom is prepared incorrectly.
The register can then be divided into two subsets: correctly prepared atoms, which occur with probability $1-\eta$, and incorrectly prepared atoms, which occur with probability $\eta$.
We denote here by $P_g^{\rm meas}$ and $P_r^{\rm meas}$ the measured ground- and Rydberg-state populations, respectively, and by $P_g^{\rm corr}$ and $P_r^{\rm corr}$ the corresponding populations at the end of the evolution for correctly prepared atoms. Neglecting measurement errors, the measured populations are given by
\begin{equation}
\begin{split}
    P_g^{\rm meas} &= \eta_g + (1-\eta) P_g^{\rm corr} \, ,\\
    P_r^{\rm meas} &= \eta_r + (1-\eta) P_r^{\rm corr} \, ,
\end{split}
\end{equation}
where $\eta_g$ and $\eta_r$ denote the fractions of incorrectly prepared atoms that are ultimately identified as being in the ground and Rydberg states, respectively, and satisfy $\eta_g+\eta_r=\eta$.
The limiting case $\eta_g=\eta$ corresponds to all incorrectly prepared atoms occupying other states within the $\ket{5s_{1/2}}$ manifold, which get recaptured during the measurement stage and thus identified as being in $\ket{g}$. Conversely, the case $\eta_r=\eta$ describes the situation in which all incorrectly prepared atoms fail to be recaptured and therefore contribute to the measured population assigned to $\ket{r}$.
It worth noting that, in the case of single-qubit experiments, where interactions are not present and hence an incorrectly prepared atom does not alter the evolution of a multiqubit state, state-preparation errors can be recast as measurement errors by appropriate redefinition of error rates (cf. Sec.~\ref{sec:meas-errors} below).

\subsubsection{Measurement errors}\label{sec:meas-errors}

Measurement errors account for the incorrect assignment of an atom in the Rydberg state to the ground state, or vice versa, during the measurement process. These errors can be modeled by introducing two parameters: the false-positive probability $\FP$ and the false-negative probability $\FN$. We identify $\ket{g}$ as the ``negative'' state and $\ket{r}$ as the ``positive'' state. Accordingly, $\FN$ is the probability of detecting an atom in $\ket{r}$ as being in $\ket{g}$, whereas $\FP$ is the probability of detecting an atom in $\ket{g}$ as being in $\ket{r}$.

Let $P_g^{\rm true}$ and $P_r^{\rm true}$ denote the true probabilities of the states $\ket{g}$ and $\ket{r}$, respectively, with
\begin{equation}
    P_r^{\rm true}=\langle n^z\rangle,
    \qquad
    P_g^{\rm true}=1-P_r^{\rm true} \, .
\end{equation}
Introducing the true-negative and true-positive probabilities,
\begin{equation}
    {\rm TN}=1-\FP,
    \qquad
    {\rm TP}=1-\FN \, ,
\end{equation}
the relationship between the true and experimentally observed probabilities is
\begin{equation}
    \left(
    \begin{matrix}
        {\rm TN} & \FN \\
        \FP & {\rm TP}
    \end{matrix}
    \right)
    \left(
    \begin{matrix}
        P_g^{\rm true} \\
        P_r^{\rm true}
    \end{matrix}
    \right)
    =
    \left(
    \begin{matrix}
        {\rm TN}P_g^{\rm true}+\FN P_r^{\rm true} \\
        {\rm TP}P_r^{\rm true}+\FP P_g^{\rm true}
    \end{matrix}
    \right) \, .
\end{equation}
The matrix on the left-hand side is referred to as the confusion matrix, whose columns each sum to one. The vector on the left-hand side contains the true physical probabilities (i.e.~it is the diagonal of the true density matrix), whereas the vector on the right-hand side contains the experimentally observed probabilities.

For a register of $N$ qubits, we can assume that detection errors act independently on each qubit: this is a good approximation for neutral atoms, since the dominant measurement error mechanisms are single-atom processes (see more details below), and the atoms are separated by several micrometres, so crosstalk in the imaging is small. The full confusion matrix then factorizes into the $N$-fold Kronecker product of single-qubit confusion matrices:
\begin{equation}
    \left(
    \begin{matrix}
        {\rm TN} & \FN \\
        \FP & {\rm TP}
    \end{matrix}
    \right)^{\otimes N}
    \left(
    \begin{matrix}
        P_{ggg\dots}^{\rm true} \\
        \vdots \\
        P_{rrr\dots}^{\rm true}
    \end{matrix}
    \right) \, .
    \label{eq:Nconfmatrix}
\end{equation}

False-positive and false-negative errors have the effect to shift the observed Rydberg-state population. In particular, a false-negative error effectively transfers part of the observed population from $\ket{r}$ to $\ket{g}$, whereas a false-positive error produces the opposite effect.
For a single-qubit experiment in which $\langle n^z\rangle$ is measured as a function of time, false negatives shift the observed curve downward most strongly when $\langle n^z\rangle$ is close to $1$. Conversely, false positives shift the curve upward most strongly when $\langle n^z\rangle$ is close to $0$. More explicitly, the observed Rydberg-state population is
\begin{equation}
\begin{split}
    P_r^{\rm obs}
    &=
    (1-\FN)P_r^{\rm true}+\FP(1-P_r^{\rm true})\\
    &=
    \FP+(1-\FP-\FN)P_r^{\rm true} \, .
\end{split}
\end{equation}

In the neutral-atom simulator considered here, measurements are performed by recapturing atoms in $\ket{g}$ and subsequently applying fluorescence imaging (cf. Sec.~\ref{sec:hardware}); the false-negative and false-positive probabilities are therefore not generally equal. A false-negative event may occur, for example, when an atom initially in $\ket{r}$ decays to $\ket{g}$ during the measurement procedure and is subsequently recaptured by the optical tweezers.
A false-positive event may instead occur when an atom in $\ket{g}$ is lost before (or during) recapture and is consequently identified as being in $\ket{r}$, or if it gets excited during measurement. Because the physical processes underlying these two types of errors differ, their probabilities (and hence $\FN$ and $\FP$) usually do not coincide. In particular, the Fresnel device typically exhibits a larger value of $\FN$ than of $\FP$, since both the loss and spontaneous-excitation probabilities are lower than the decay one.

\subsection{Coherent errors}\label{sec:coherent}
We analyze in this Section those errors referred to as {\em coherent}, since they enter directly into the Hamiltonian and thus do not alter the unitarity of the evolution at the single-shot level. However, we remark that, when the nature of these errors is stochastic, non-unitary effects arise once averaging over several realizations, as is commonly done when repeating the same experiment many times to collect shot statistics and reconstruct observables.
In particular, in the case these errors are affecting the driving fields, they may take the form of a constant offset, a multiplicative calibration error, a time-dependent perturbation (e.g. random fluctuations) or a position-dependent distortion. In the following, we provide a brief review of (ideal) unitary dynamics, and analyze in detail how the possible errors alter it.

\subsubsection{Unitary dynamics}\label{sec:unitary}

For simplicity, let us first consider the dynamics of a single atom. Under a constant, i.e.~time-independent, drive with Rabi frequency $\Omega$ and detuning $\delta$, the Schr\"odinger equation can be solved explicitly. Assuming the initial condition $\ket{\psi(0)}=\ket{g}$, the atom undergoes Rabi oscillations according to
\begin{equation}
    \ket{\psi(t)}
    =
    \left(
        \cos\frac{\widetilde{\Omega}t}{2}
        -i\frac{\delta}{\widetilde{\Omega}}
        \sin\frac{\widetilde{\Omega}t}{2}
    \right)\ket{g}
    -i\frac{\Omega}{\widetilde{\Omega}}
    \sin\frac{\widetilde{\Omega}t}{2}\ket{r} \, ,
    \label{eq:Rabi-oscillations-detuned}
\end{equation}
where
\begin{equation}
    \widetilde{\Omega}^2=\Omega^2+\delta^2 \, .
\end{equation}
In the resonant case, $\delta=0$, Eq.~\eqref{eq:Rabi-oscillations-detuned} reduces to
\begin{equation}
    \ket{\psi(t)}
    =
    \cos\frac{\Omega t}{2}\ket{g}
    -i\sin\frac{\Omega t}{2}\ket{r} \, .
    \label{eq:Rabi-oscillations}
\end{equation}

The frequency of the oscillations between $\ket{g}$ and $\ket{r}$ is therefore primarily determined by $\Omega$. Introducing a nonzero detuning has two effects, as can be seen by comparing Eq.~\eqref{eq:Rabi-oscillations-detuned} with its resonant counterpart, Eq.~\eqref{eq:Rabi-oscillations}. First, it changes the oscillation frequency from $\Omega$ to the generalized Rabi frequency
$\widetilde{\Omega}=\sqrt{\Omega^2+\delta^2}$. Second, it limits the maximum Rydberg-state population to
\begin{equation}
    P_r^{\rm max}=\frac{\Omega^2}{\Omega^2+\delta^2} \, .
\end{equation}
Geometrically, the detuned Hamiltonian generates a Bloch-sphere rotation about the effective field $\boldsymbol{\Omega}_{\rm eff}=(\Omega,0,-\delta)$, whose magnitude is $\widetilde{\Omega}=\sqrt{\Omega^2+\delta^2}$. Relative to the resonant rotation axis, the effective axis is tilted in the $xz$ plane by an angle equal to $\arctan(-\delta/\Omega)$.

Analogous considerations can be extended to initial states satisfying $\ket{\psi(0)}\neq\ket{g}$ and to time-dependent driving fields.
The following observations are particularly useful for the subsequent analysis of coherent errors:
\begin{enumerate*}[label=(\roman*)]
    \item detuning noise applied to a resonant drive increases the generalized Rabi frequency, although it simultaneously reduces the maximum achievable Rydberg-state population;
    \item detuning noise applied to an already detuned drive can either increase or decrease the generalized Rabi frequency, depending on the sign of the perturbation relative to the nominal detuning; and
    \item for a resonant drive, ignoring other effects, oscillations slower than the nominal prediction indicate that the drive amplitude is not fully delivered, namely that the amplitude experienced by the system is lower than the one requested by the user.
\end{enumerate*}

\subsubsection{Laser waist}
The laser-waist effect arises from the finite transverse extent of the laser beam. The transverse intensity profile can typically be approximated by a Gaussian function, whose characteristic width is described by the beam waist $w$. Consequently, atoms located farther from the laser propagation axis experience a lower laser intensity and, in the present setting, a smaller effective Rabi frequency. We model this spatial dependence as
\begin{equation}
    \Omega(r)=\Omega_0\exp\left(-\frac{r^2}{w^2}\right) \, ,
\end{equation}
where $r$ denotes the distance from the propagation axis and $\Omega_0$ is the (generally time-dependent) amplitude requested by the user.

In the Fresnel device considered in this work, the atoms are arranged in the $xy$ plane and the laser propagates along the $y$ direction. The transverse distance from the optical axis, identified with the $y$ axis, is therefore given by the $x$ coordinate of each atom. The time- and position-dependent Rabi frequency experienced by an atom is thus
\begin{equation}
    \Omega(t,x)
    =
    \Omega_0(t)\exp\left(-\frac{x^2}{w^2}\right) \, ,
\end{equation}
where $\Omega_0(t)$ here denotes the nominal laser amplitude requested by the user.

Since, in the present work, we performed many parallel experiments on (clusters of) atoms located at different positions within the accessible trapping region, a significant fraction of them experienced a driving field with a Rabi frequency $\Omega$ lower than the nominally requested value. Averaging over the positions of the atomic clusters therefore reveals two distinct consequences of the finite laser waist. First, the population dynamics exhibit slower oscillations, as the effective Rabi frequency experienced by the ensemble is, on average, smaller than the target value. Second, the oscillations become damped, since the measured signal results from averaging oscillatory responses characterized by different oscillation periods.

\subsubsection{Doppler effect and quasi-static detuning noise}
Another source of noise in the driving fields experienced by the atoms is due to their residual thermal motion at finite temperature. In particular, an atomic ensemble at temperature $\mathfrak{t}$ has a thermal velocity distribution described by the Maxwell--Boltzmann distribution. At the equilibrium, each component of the atomic velocity is normally distributed with zero mean and standard deviation
\begin{equation}
\Delta v = \sqrt{\frac{k_{\mathrm B}\mathfrak{t}}{m}} \, ,
\end{equation}
where $k_{\mathrm B}$ is the Boltzmann constant and $m$ is the atomic mass.
Because of the Doppler effect, the motion of an atom induces a velocity-dependent shift of the laser frequency in the atomic rest frame, which can be effectively described as an additional contribution to the detuning. For an atom with velocity $\vec{v}$, the corresponding Doppler shift is
\begin{equation}
\delta_{\rm qs} = \vec{k}_{\rm eff}\cdot\vec{v} \, , \label{eq:doppler_shift}
\end{equation}
where, $\vec{k}_{\rm eff}$ is the effective wave vector of the driven transition.
If the atomic velocity remains approximately constant over the duration of a single experimental sequence, but varies independently between different realizations, the Doppler shift can be modeled as a quasi-static random detuning offset. In particular, the projection of the velocity along $\vec{k}_{\rm eff}$ being Gaussian distributed as well, the nominal time-dependent detuning is replaced by
\begin{equation}
\delta(t) \longmapsto \delta(t)+\delta_{\rm qs} \, ,
\end{equation}
where $\delta_{\rm qs}$ is sampled for each shot from the Gaussian distribution
\begin{equation}
\delta_{\rm qs}\sim \mathcal{N}(\delta_0,\sigma_\delta^2) \, ,
\end{equation}
with 
\begin{equation}
\sigma_\delta = \lvert\vec{k}_{\rm eff}\rvert \sqrt{\frac{k_{\mathrm B}\mathfrak{t}}{m}}
\end{equation}
and $\delta_0$ accounting for the presence of additional systematic drifts or offsets.

For each fixed realization of quasi-static detuning noise, the evolution remains unitary since $\delta_{\rm qs}$ enters as a static contribution to the Hamiltonian. Experimental observables, however, are obtained by averaging over many realizations, each characterized by a different value of $\delta_{\rm qs}$, and although every individual realization is governed by unitary dynamics, the ensemble-averaged evolution is generally described by an effectively non-unitary channel.
This inhomogeneous averaging leads to a dephasing-like mechanism and a progressive reduction of the observed coherent contrast, but with an important distinction depending on the presence or not of a systematic nonzero detuning.

In formulae, the loss of oscillation contrast can be directly calculated. Let $P_r^{\rm avg}$ denote the probability of detecting an atom in $\ket{r}$ after averaging over several realizations of $\delta_{\rm qs}$.
In the resonant case $\delta=\delta_0=0$ and under the assumption $\sigma_\delta\ll\Omega$, the Rabi oscillations get damped with an algebraic envelope
\begin{equation}
    P_r^{\rm avg}(t) \approx \frac12 \left[\vphantom{\frac12}1- C(t)\cos\big(\Omega t + \varphi(t)\big)\right] \, ,
\end{equation}
with
\begin{equation}
    C(t) = \left[1+\left(\frac{\sigma_\delta^2t}{\Omega}\right)^2\right]^{-1/4} \, ,
\end{equation}
which approaches $P_r^{\rm avg}=1/2$ as $t^{-1/2}$.
On the other hand, for a non-zero detuning $\delta$, holding both $\sigma_\delta\ll\Omega$ and $\sigma_\delta\ll\lvert\delta\rvert$, the average population gets Gaussianly damped
\begin{equation}
    P_r^{\rm avg}(t) \approx
    \frac12\frac{\Omega^2}{\Omega^2+\delta^2}
    \left[
    1-C(t)\cos\left(t\sqrt{\Omega^2+\delta^2}\right) \right] \, ,
\end{equation}
with
\begin{equation}
    C(t) = \exp\left[-\frac12\left(\frac{\delta\sigma_\delta t}{\sqrt{\Omega^2+\delta^2}}\right)^2\right] \, ,
\end{equation}
which has asymptotic population value $P_r^{\rm avg}=\Omega^2/[2(\Omega^2+\delta^2)]$.
We further note that, in the detuned case, the average frequency of Rabi oscillations keeps being $\sqrt{\Omega^2+\delta^2}$ regardless the presence of noise, while, in the resonant case, the average frequency of Rabi oscillations gets modified by an amount depending on $\sigma_\delta^2$ (cf. Eq.~\eqref{eq:quasi-static-resonant-envelope-phase}).
The different behavior comes from the interplay of resonant (or detuned) dynamics and noise magnitude, and detailed computations can be found in App.~\ref{app:quasi-static}.

It is important to remark that residual thermal motion is not the only physical mechanism that can generate quasi-static detuning noise. Other sufficiently slow fluctuations may contribute to the effective values of $\delta_0$ and $\sigma_\delta$ and can be included in the same model, for instance low-frequency laser phase noise, and fluctuations induced by the control electronics or the surrounding environment.

Let us conclude this Section with some computations using realistic experimental values. Fresnel is based on Rubidium atoms, whose mass is $m=1.45\times10^{-25}$~kg, and is able to cool them down to a few tens of micro-Kelvin degrees; recalling the value of the Boltzmann constant $k_B=1.38\times10^{-23}$~J/K and using the value $\lvert\vec{k}_{\rm eff}\rvert=8.7$~$\mu$m$^{-1}$ from Ref.~\cite{deLsleuc2018}, we obtain a value of $\sigma_\delta\approx0.38$~rad/$\mu$s at $\mathfrak{t}=20$~$\mu$K, which increases to $\sigma_\delta\approx0.60$~rad/$\mu$s at $\mathfrak{t}=50$~$\mu$K.

\subsection{Open-system errors}\label{sec:opensys}
Finally, we analyze in this Section those errors that arise due to interaction of the system with the environment. In particular, under the assumption of Markovian dynamics, we include these errors as jump operators $L_k$ in the Lindblad master equation
\begin{equation}
    \dot \rho = -i\:[H, \rho] + \sum_k \left( L_k \rho L_k^\dag - \frac{1}{2} \{L_k^\dag L_k, \rho\} \right) \, ,
\end{equation}
where $\rho$ denotes the density matrix describing the system, $[A,B]=AB-BA$ the commutator, and $\{A,B\}=AB+BA$ the anticommutator.
In the following, we will analyze in detail the effect on dynamics of dephasing, i.e.~the loss of phase coherence, and decay, i.e.~the spontaneous transition from the state $\ket{1}$ to the state $\ket{0}$.

\subsubsection{Dephasing}\label{sec:dephasing}
Markovian dephasing is an error mechanism that accounts for the loss of phase coherence between states in a coherent superposition, which we model with one jump operator per atom,
\begin{equation}
    L_j=\ketbra[j]{r}{r} \, ,
\end{equation}
or, equivalently up to an additive identity contribution and a multiplicative factor, by $L_j=\sigma_j^z$, with $j=1,\dots,N$.

In the resonant case, the single-qubit Bloch equations yield a differential equation analogous to that of a damped harmonic oscillator (see App.~\ref{app:dephasing} for detailed computations). For the initial state $\ket{\psi(0)}=\ket{g}$, which is the initial state used in our experiments, the solution reads
\begin{equation}
    s_z(t)
    =
    e^{-\frac{\gamma_\phi}{2}t}
    \cos\left(
        \frac{\sqrt{\Omega^2-\gamma_\phi^2}}{2}t
    \right) \, ,
\end{equation}
where $s_z=\langle\sigma^z\rangle$ denotes the expectation value of the spin component along the $z$ axis and $\gamma_\phi$ is the dephasing rate.

This expression suggests that dephasing can also reduce the apparent oscillation frequency, in addition to a lower effective value of $\Omega$ or a detuning bias (cf. Sec.~\ref{sec:unitary}). However, the magnitude of this effect can be estimated quantitatively. For an evolution time of up to $T_{\rm max}=6$~$\mu$s, an average Rabi frequency $\Omega\approx5$~rad/$\mu$s, and a dephasing rate $\gamma_\phi\approx0.6$~rad/$\mu$s, using $L_j=n_j^z$ as the jump operator, Eq.~\eqref{eq:phase-dephasing} gives an accumulated phase difference relative to ideal Rabi oscillations of approximately
\begin{equation}
    \Delta\varphi\approx0.055~{\rm rad}\approx0.017\pi~{\rm rad} \, .
\end{equation}
This correction is comparatively small. For the same experimental parameters, a $2\%$ reduction in the nominal Rabi frequency would instead produce a phase shift of approximately $\Delta\varphi=0.6~{\rm rad}$, which is about one order of magnitude larger. We therefore expect an observable reduction in the Rabi-oscillation frequency to be caused primarily by amplitude noise rather than by dephasing.

In contrast, the damping of the oscillations induced by dephasing cannot generally be neglected. In the limit $\gamma_\phi t\gg1$, the single-qubit dynamics approaches $s_z=0$, corresponding to equal populations of the ground and Rydberg states, and therefore to a $50\%$ probability of detecting the atom in $\ket{r}$.
The analytical solution for detuned case cannot be obtained simply by tilting the corresponding resonant solution. The coherent dynamics, indeed, rotates the state about a different Bloch-sphere axis, whereas dephasing continues to act along the $z$ direction. Their combined action therefore produces more complex dynamics. Nevertheless, empirical considerations are similar to the ones for the resonant case.

\subsubsection{Decay}\label{sec:decay}
This effect describes the spontaneous decay of an electron from the Rydberg state to the ground state. In computational terms, it corresponds to a population transfer from $\ket{r}$ to $\ket{g}$ and is represented, independently for each atom, by the jump operator
\begin{equation}
    L_j = \ketbra[j]{g}{r} = \sigma^-_j \, ,
\end{equation}
where $j=1,\dots,N$.
Considering a single atom starting in $\ket{\psi(0)}=\ket{g}$ under the effect of a resonant drive,
the integration of the Lindblad master equation
yields a dynamics similar to what observed in the presence of dephasing, but with an important distinction: also in this case the error rate plays the role of damping the harmonic oscillations generated by a resonant drive, but at the same time it also acts as a force pushing the population towards the ground state.
Further computations can be found in App.~\ref{app:decay}, and the solution reads
\begin{equation}
    s_z(t) = A + B e^{-\frac{3\Gamma}{4} t} \cos(\Omega_\Gamma t - C) \, ,
\end{equation}
where $s_z=\langle\sigma^z\rangle$, $\Omega_\Gamma=\sqrt{\Omega^2-\Gamma^2/16}$, and $A,B,C$ are quantities depending on $\Omega,\Gamma$.

Similarly to the dephasing case, also spontaneous decay in principle modifies the apparent Rabi-oscillation frequency. In practice, however, its effect is very limited: Rydberg-atom experiments are typically characterized by long Rydberg-state lifetimes, corresponding to decay rates of approximately $\Gamma \approx 0.01$~$\mu$s$^{-1}$.
Consequently, once again, any observed shift in the Rabi-oscillation frequency is expected to arise primarily from amplitude noise rather than spontaneous decay.

Also in this case, the analytical solution for a detuned drive cannot be obtained simply by tilting the resonant one, since the decay remains fixed along the $z$-axis, which no longer coincides with the axis about which the system rotates, but empirical considerations still hold.

\section{Experiments performed}\label{sec:experiments}

In the setting of a neutral-atom quantum simulator, the input data are given by the time-dependent control waveforms and the coordinates of the atoms in the register, whereas the output data consist of projective measurement outcomes in the computational basis.
The characterization task is then complicated by the fact that different noise mechanisms typical of this hardware may produce similar signatures in the accessible observables, as discussed in Sec.~\ref{sec:noise}.
Distinguishing these mechanisms becomes particularly difficult when each experimental protocol is sampled at only one final evolution time. Indeed, a single measured population generally provides little information to distinguish errors accumulated during the dynamics from state-preparation and measurement errors; for instance, an appropriate choice of the readout-error probabilities for a single-qubit case may reproduce any observed final population independently of the preceding dynamics. This lack of identifiability motivates the design of experiments that not only reveal deviations from the ideal dynamics, but also distinguish, as far as possible, the characteristic signatures of different noise sources.
We remark that actually, in principle, a sufficiently diverse dataset composed only of final-time measurements may still contain substantial information about the underlying noise processes, especially when combined with suitable fine postprocessing methods~\cite{Krastanov2019,Canonici2024}.
In the present work, however, we find time-resolved experiments to be particularly useful for device characterization, since, besides providing greater interpretability, their temporal structure additionally facilitates the construction of a physically motivated ansatz for the effective model.

\begin{figure}
    \centering
    \includegraphics[width=0.95\linewidth]{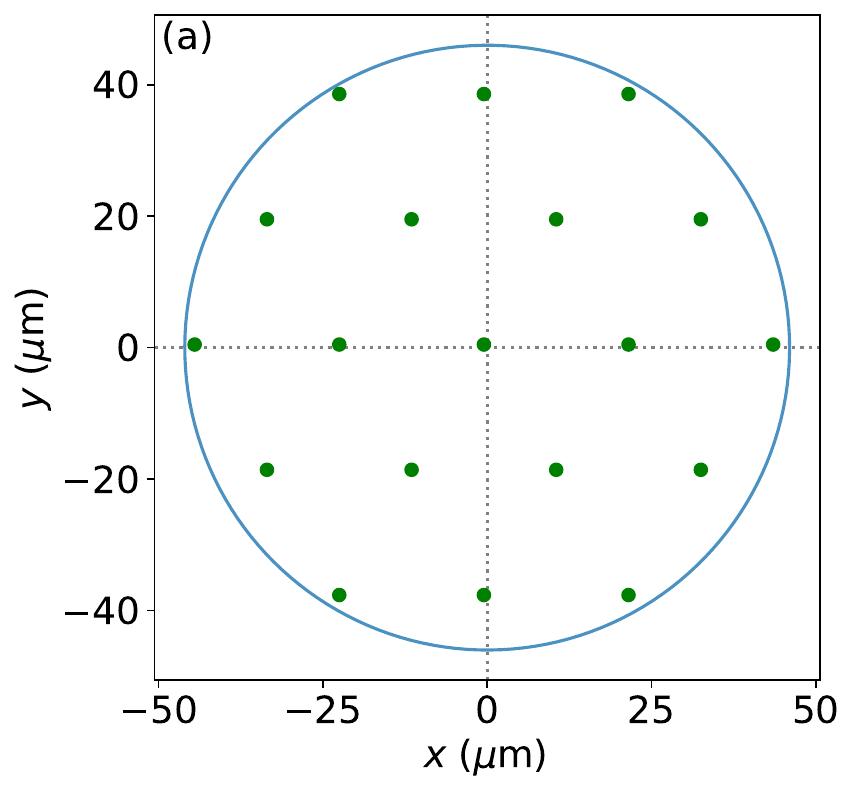}
    \includegraphics[width=0.95\linewidth]{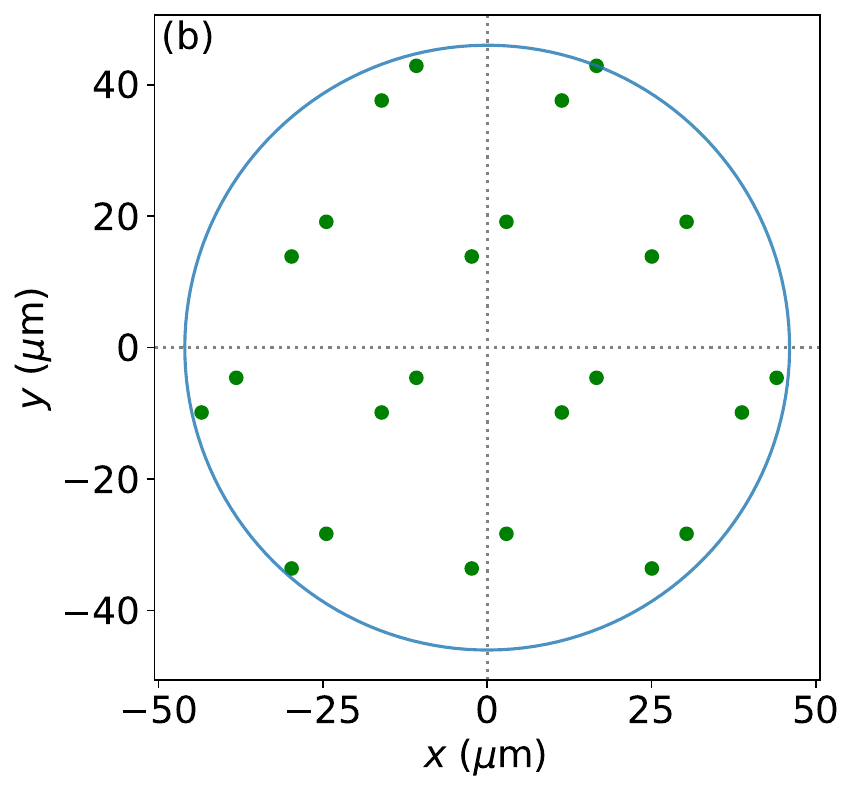}
    \cprotect\caption{Registers used for performing experiments investigating (a) single-atom and (b) two-atoms dynamics. Atoms are represented by green dots, while the circle denotes the region in which atoms are allowed to be placed; in (b), the intra-pair distance has been changed from experiment to experiment in order to investigate different interaction regimes. The (pairs of) atoms are placed sufficiently far away to approximate their dynamics as independent.}
    \label{fig:dyn-registers}
\end{figure}

For our characterization protocol, we investigated two main kinds of experiments: single-qubit and two-qubit gates. 
We decided not to investigate multiqubit gates involving more than two atoms, both because their dynamics becomes increasingly intricate and difficult to interpret and, more importantly, because the Rydberg Hamiltonian contains at most two-body interactions. Given also the further intrinsic equivalence of all atoms, we expected that studying gates involving up to two qubits would be sufficient to identify the dominant noise contributions to the dynamics.
For both kinds of experiments investigated, we performed experiments by putting far away (pairs of) atoms in the allowed region, thus parallelizing the execution of experiments over many independent subsystems (see Fig.~\ref{fig:dyn-registers}). We have seen that this procedure has not introduced biases in the data we present here due to the presence of residual interactions: indeed, these are in the order of $10^{-3}$~rad/$\mu$s can be considered negligible.

In particular, we have executed the following experiments:
single-atom experiments with resonant drive (i.e.~$\delta(t)=0$);
single-atom experiments with detuned drive;
two-atoms experiments with amplitude-only drive, potentially changing the interaction strength (i.e.~the interatomic distance); and
two-atoms experiments with detuned drive.

\begin{figure}
    \centering
    \includegraphics[width=0.95\linewidth]{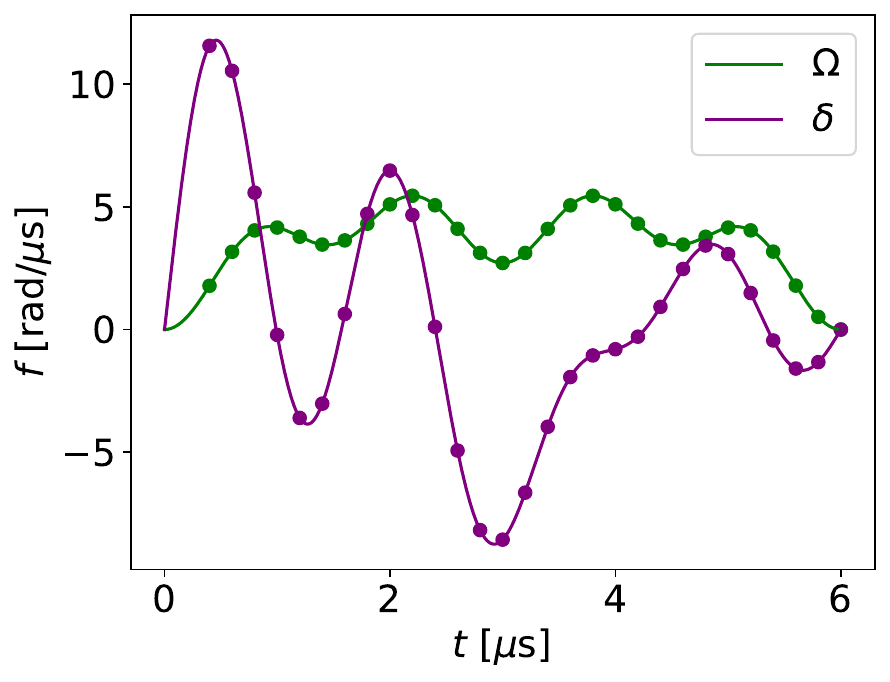}
    \cprotect\caption{Example of pulses executed for investigating the dynamics of one- and two-atom experiments, corresponding in this case to \verb|dyn_1|. Dots correspond to times at which measurements have been performed.}
    \label{fig:example-pulses}
\end{figure}

In each case the time-dependent field has been obtained as a random pulse consisting in the superposition of (a few) sinusoidal or cosinusoidal waves with random magnitude and low frequency, namely
\begin{equation}
\begin{split}
    \Omega(t)&=\sum_{k=1}^{K-1} \frac{c_k}2\left[1-\cos\left(\frac{2\pi k} T t\right)\right]\,,\quad c_k\geq0\:\forall\:k\\
    \delta(t)&=\sum_{k=1}^{2(K-1)} c_k\sin\left(\frac{\pi k}T t\right)
\end{split}
\end{equation}
with $K\in[2,6]$ appropriately chosen depending on $T$ in order get low-bandwidth pulses, and $c_k$ randomly sampled.
Constraints on maximum and minimum power of the fields have been included by rescaling the sampled pulse shapes as $f(t) \mapsto A f(t)/ \max_t \lvert f(t) \rvert$, with $A$ uniformly sampled in the interval $[f_{\rm min}, f_{\rm max}]$, and $f$ denoting either $\Omega$ or $\delta$.
A representative example of the pulse shapes implemented can be found in Fig.~\ref{fig:example-pulses}.

As anticipated, for experiments presented here, we followed the real-time dynamics of atoms' population. In order to do so, we truncated the pulse at intermediate times and we performed the measurements of the current atoms' population, repeating the procedure multiple times in order to reconstruct population probabilities. 

In the following, we will report detailed presentation of the experiments performed, grouped by kind. A brief summary of them can be found in Tab.~\ref{tab:experiments}.
Overall, we performed 6030 shots for single-qubit experiments and 8150 shots for two-qubit experiments, for a total amount of 14180 shots.

\begin{table}
    \centering
    \begin{tabular}{|c|c|c|c|c|}
         \hline
         id&n. qubits&$T$~[ns]&$\delta(t)\neq0$&$U$~[rad/$\mu$s]\\\hline
         \verb|dyn_0|&1&6000&false&0\\\hline
         \verb|dyn_1|&1&6000&true&0\\\hline
         \verb|dyn_2|&1&6000&false&0\\\hline
         \verb|dyn_3|&1&6000&true&0\\\hline
         \verb|dyn_4|&1&6000&true&0\\\hline
         \verb|dyn_5|&2&6000&true&16\\\hline
         \verb|dyn_6|&2&6000&false&7.47\\\hline
         \verb|dyn_7|&2&6000&false&55.41\\\hline
         \verb|dyn_8|&2&4000&false&[11.48, 4.86, 2.29, 1.17]\\\hline
         \verb|dyn_9|&2&4000&false&[11.48, 4.86, 2.29, 1.17]\\\hline
         \verb|dyn_10|&2&4000&false&[11.48, 4.86, 2.29, 1.17]\\\hline
    \end{tabular}
    \caption{Overview of experiments performed and associated relevant quantities.}
    \label{tab:experiments}
\end{table}

\paragraph{Real time dynamics of single-atom experiments with resonant drive}

These datasets consist in the following experiments performed over 19 distant atoms in parallel (see Fig.~\ref{fig:dyn-registers}(a)):
\begin{itemize}[label={}, leftmargin=0pt, nosep]
    \item \verb|dyn_0|: repeated 60 times for each sampled point in the experimental time span $[400,6000]$~ns with a sampling rate of 200~ns;
    \item \verb|dyn_2|: repeated 50 times for each sampled point in the experimental time span $[600,6000]$~ns with a sampling rate of 300~ns.
\end{itemize}

\paragraph{Real time dynamics of single-atom experiments with detuned drive}
These datasets consist in the following experiments performed over 19 distant atoms in parallel (see Fig.~\ref{fig:dyn-registers}(a)):
\begin{itemize}[label={}, leftmargin=0pt, nosep]
    \item \verb|dyn_1|: repeated 60 times for each sampled point in the experimental time span $[400,6000]$~ns with a sampling rate of 200~ns;
    \item \verb|dyn_3|: repeated 50 times for each sampled point in the experimental time span $[800,6000]$~ns with a sampling rate of 400~ns, using the same pulse shape of \verb|dyn_1|;
    \item \verb|dyn_4|: repeated 50 times for each sampled point in the experimental time span $[600,6000]$~ns with a sampling rate of 300~ns.
\end{itemize}
We highlight the fact that experiments \verb|dyn_1| and \verb|dyn_3| have the same pulse shape but they have been performed at several weeks of time delay, with the aim to better check changes of the hardware over time.

\paragraph{Real time dynamics of two-atoms experiments with amplitude-only drive}

These datasets consist in the following experiments performed over 12 distant pairs of atoms in parallel (24 atoms in total):
\begin{itemize}[label={}, leftmargin=0pt, nosep]
    \item \verb|dyn_6|: repeated 50 times for each sampled point in the experimental time span $[600,6000]$~ns with a sampling rate of 300~ns, with interaction strength $U=5$~rad/$\mu$s ($r=7.47$~$\mu$m);
    \item \verb|dyn_7|: repeated 50 times for each sampled point in the experimental time span $[600,6000]$~ns with a sampling rate of 300~ns, with interaction strength $U=55.41$~rad/$\mu$s ($r=5$~$\mu$m);
    \item \verb|dyn_8|, \verb|dyn_9|, and \verb|dyn_10|: repeated 40 times for each sampled point in the experimental time span $[400,4000]$~ns with a sampling rate of 400~ns, with interaction strength $U\in[11.48,\allowbreak4.86,\allowbreak2.29,\allowbreak1.17]$~rad/$\mu$s (\allowbreak$r\in[6.5,\allowbreak7.5,\allowbreak8.5,\allowbreak9.5]$~$\mu$m).
\end{itemize}

\paragraph{Real time dynamics of two-atoms experiments with detuned drive}

These datasets consist in the following experiments performed over 12 distant pairs of atoms in parallel (24 atoms in total):
\begin{itemize}[label={}, leftmargin=0pt, nosep]
    \item \verb|dyn_5|: repeated 50 times for each sampled point in the experimental time span $[400,6000]$~ns with a sampling rate of 200~ns, with interaction strength $U=16$~rad/$\mu$s ($r=6.15$~$\mu$m);
\end{itemize}

\section{Investigation of experimental data}\label{sec:investigation}

Having presented an overview of the hardware and of the effects of noise in Secs.~\ref{sec:hardware} and \ref{sec:noise}, respectively, we now analyze the data obtained from the experiments described in Sec.~\ref{sec:experiments}.

As a first step, in Secs.~\ref{sec:1q-exp} and \ref{sec:2q-exp} we examine the trajectories returned by the quantum simulator and compare them with the predictions of an ideal model governed by the Rydberg Hamiltonian of Eq.~\eqref{eq:HRyd3}.
This comparison gives access to oscillation frequencies, damping envelopes, population extrema, and long-time behavior.
Taken together, these features constrain an effective noise model more tightly than measurements at a single evolution time, so that distinct insights into the system can be extracted from only a few measurements.
In this way, we identify the main noise signatures and associate them with their likely sources. For example, a control-amplitude error primarily modifies the coherent Rabi frequency, whereas Markovian dephasing or a distribution of quasi-static detunings suppresses the oscillation contrast.
Although these signatures are not, in general, fully independent, and in some cases are difficult to tell apart (e.g., both Markovian dephasing and quasi-static detuning noise damp the oscillations), their full temporal dependence improves parameter identifiability and helps separate the contributions of different error mechanisms.

Building on these observations, in Sec.~\ref{sec:model} we develop an informed effective model of the hardware, motivating the inclusion (or exclusion) of specific noise effects.

\subsection{Single-qubit experiments}\label{sec:1q-exp}

We begin with single-qubit experiments, for which the effects of different noise sources are easier to isolate and interpret. To illustrate our procedure, we consider the data obtained from experiment \verb|dyn_0|. The same reasoning can be applied directly to the other single-qubit experiments.

\begin{figure}
    \centering
    \includegraphics[width=0.95\linewidth]{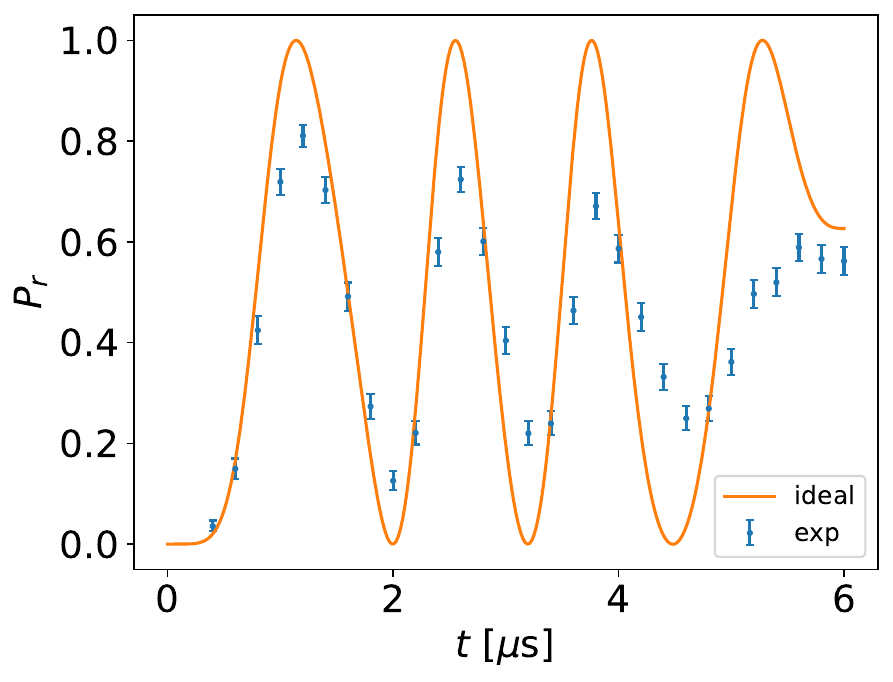}
    \cprotect\caption{Comparison between the experimental data obtained from experiment \verb|dyn_0| (averaged over atoms' positions) and the predictions of a completely ideal model. Error bars on the experimental data are obtained with the Bayesian estimation procedure described in App.~\ref{app:bayes}.}
    \label{fig:dyn_0-ideal}
\end{figure}

Fig.~\ref{fig:dyn_0-ideal} compares the experimentally measured Rydberg-state population, averaged over all atoms in the register, with the dynamics predicted under ideal conditions. Three main discrepancies are immediately apparent: (i) the observed oscillation frequency is lower than its ideal value; (ii) the oscillations are damped towards $P_r=0.5$; and (iii) the Rydberg-state population never reaches $P_r=1$ and, in particular, does not exceed approximately $P_r=0.85$. According to the discussion of noise sources in Sec.~\ref{sec:noise}, these effects may be associated, respectively, with (i) laser-amplitude noise, including the laser-waist effect; (ii) dephasing and detuning noise, the latter originating from either laser noise or the Doppler effect; and (iii) strong dephasing, detuning noise, and SPAM errors.

\begin{figure}
    \centering
    \includegraphics[width=0.95\linewidth]{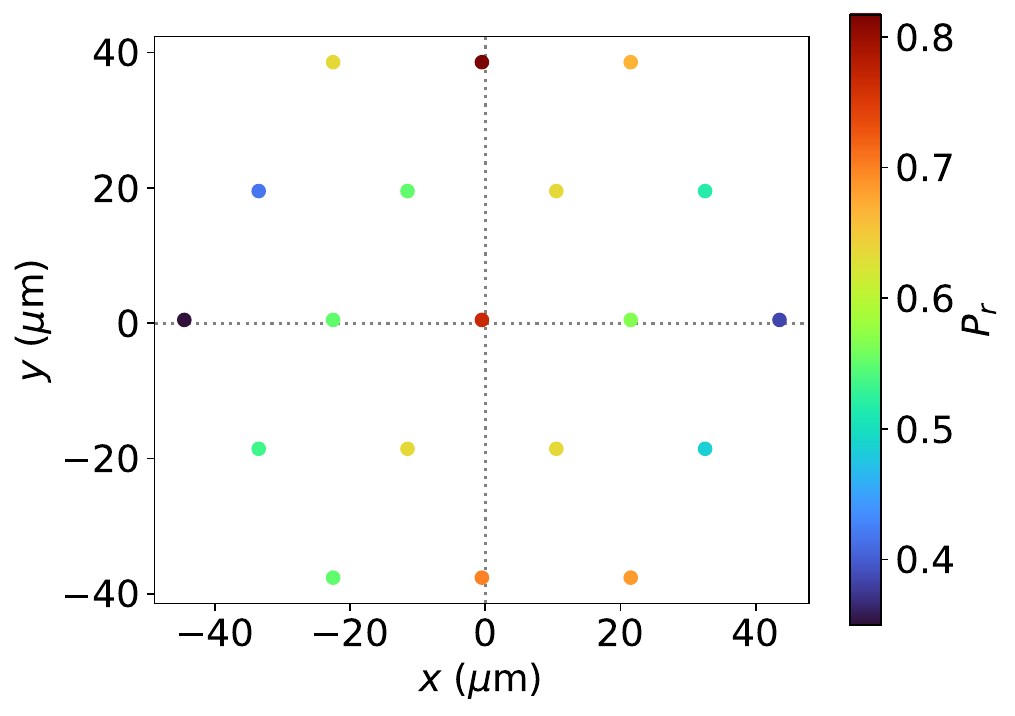}
    \cprotect\caption{Spatial variation of the excitation probability induced by the finite laser waist. The atoms are sufficiently far apart to be treated as independent. Since the laser propagates along the $y$ direction, the drive amplitude varies with the $x$ coordinate of each atom. The data are obtained from the measurement outcomes of experiment \verb|dyn_0| at $t=2400$~ns.}
    \label{fig:register-waist}
\end{figure}

\begin{figure*}
    \centering
    \includegraphics[width=0.95\linewidth]{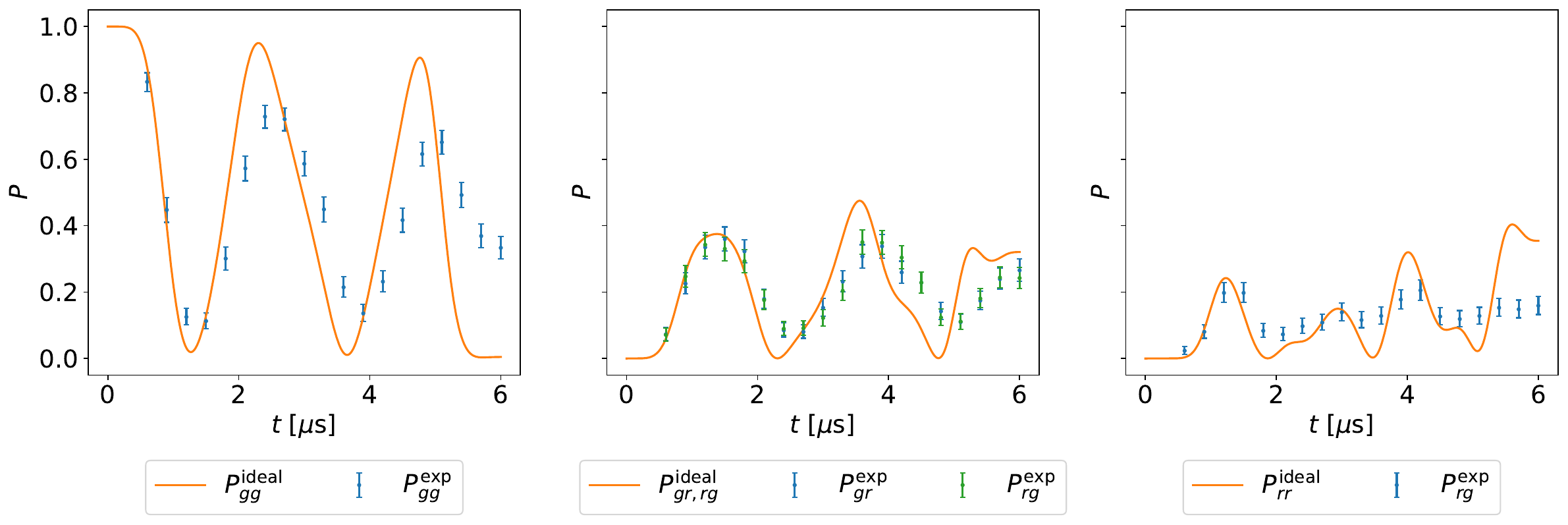}
    \cprotect\caption{Comparison between the experimental data obtained from experiment \verb|dyn_6| (averaged over pairs of atoms) and the predictions of a completely ideal model. Error bars on the experimental data are obtained with the Bayesian estimation procedure described in App.~\ref{app:bayes}.}
    \label{fig:dyn_6-ideal}
\end{figure*}

The presence of a finite laser waist can be assessed by examining the dynamics of individual atoms rather than the population averaged over the entire register. Fig.~\ref{fig:register-waist} provides a representative snapshot at $t=2400$~ns of the experiment \verb|dyn_0|. The figure reveals substantial differences between the Rydberg-state populations of atoms located close to the $y$ axis and those located farther away from it.
This observation indicates that different atoms experience different effective Rabi frequencies, with, in particular, atoms located farther from the $y$ axis experiencing a lower drive amplitude. We can therefore claim that the laser-waist effect contributes to the reduction in the oscillation frequency observed in Fig.~\ref{fig:dyn_0-ideal} after averaging over the entire register.

\subsection{Two-qubit experiments}\label{sec:2q-exp}

We next investigate datasets involving pairs of interacting atoms. As in the single-qubit analysis, we use experiment \verb|dyn_6| as a representative example in this Section, and the same reasoning can be applied to the other two-qubit experiments.

Fig.~\ref{fig:dyn_6-ideal} compares the experimentally measured populations of the two-qubit basis states with the corresponding ideal dynamics, after averaging over pairs in the register. Specifically, Fig.~\ref{fig:dyn_6-ideal}(a) shows the population of $\ket{gg}$, Fig.~\ref{fig:dyn_6-ideal}(b) shows the populations of $\ket{gr}$ and $\ket{rg}$, and Fig.~\ref{fig:dyn_6-ideal}(c) shows the population of $\ket{rr}$.

As in the single-qubit case, three main discrepancies can be identified: (i) the observed oscillation frequency is lower than the ideal one; (ii) the oscillations are damped; (iii) none of the state populations reaches either $0$ or $1$ during the dynamics; and (iv) oscillations of the $\ket{rr}$ state appear relatively more damped than for other states. These discrepancies can be associated, respectively, with (i) laser-amplitude noise and (ii)--(iv) dephasing, detuning noise, and SPAM errors. As in the single-qubit case, the laser-waist effect can account for part of the observed reduction in the oscillation frequency.

\subsection{Model ansatz construction}\label{sec:model}
Building on the comparison between the experimental data and the ideal dynamics carried out in the previous Secs.~\ref{sec:1q-exp} and \ref{sec:2q-exp}, we now construct an effective model describing Fresnel.
In principle, each of the noise sources presented in Sec.~\ref{sec:noise} may contribute to the observed deviations from the ideal dynamics. We nevertheless omit state-preparation errors from our model. Indeed, in neutral-atom platforms, optical pumping into $\ket{g}$ is typically highly efficient, and the register is verified to be defect-free (i.e.~no atom is missing) before each shot~\cite{deLsleuc2018,Bernien2017}.

Other noise sources are instead included in our model, and in particular we have:
\begin{enumerate*}[label=(\roman*)]
    \item \emph{measurement errors}: false-positive and false-negative detection probabilities, denoted by $\FP$ and $\FN$, respectively;
    \item \emph{coherent errors}: a global multiplicative correction $m_\Omega$ to the Rabi frequency, a finite laser waist $w$, and a quasi-static detuning distribution $\delta_{\rm qs}\sim\mathcal{N}(\delta_0,\sigma_\delta^2)$;
    \item \emph{open-system noise}: dephasing, described by the jump operators $n_j^z$ with dephasing rate $\gamma_\phi$, and decay, described by the jump operators $\sigma_j^-=\ketbra[j]{g}{r}$ with decay rate $\Gamma$.
\end{enumerate*}

Before proceeding, we note that an additional noise source within our framework could affect the experimental data. Recalling that the time-resolved data were obtained by truncating the pulses at intermediate times, since the laser amplitude has a finite modulation bandwidth, an abrupt truncation of the pulse sequence may leave a residual drive while the laser switches off. This effect could be modeled, for instance, by appending an $R_X(\theta_\Omega)$ global single-qubit rotation at the end of the evolution, with $\theta_\Omega$ proportional to the amplitude at the truncation time, or by accounting for more complex distortions of the input fields~\cite{Singh2023}. We omit this effect as well, since some tests performed indicated that it should be negligible.

The corresponding simulation pipeline for the effective model consists then of the following steps:
\begin{enumerate}
    \item Consider the requested control pulses $\Omega(t)$ and $\delta(t)$, with total duration $T$, and the register topology (in particular the intra-pair interaction $U$ for two-qubit experiments).

    \item For a given realization of the quasi-static detuning offset $\delta_{\rm qs}$, evolve the system up to time $T$ according to the Lindblad master equation
    \begin{equation}
    \begin{split}
        \dot{\rho}
        =
        &-i[H,\rho] \\
        &+
        \sum_\alpha
        \gamma_\alpha
        \left(
            L_\alpha\rho L_\alpha^\dagger
            -\frac{1}{2}
            \left\{
                L_\alpha^\dagger L_\alpha,\rho
            \right\}
        \right) \, ,
    \end{split}
    \end{equation}
    with Hamiltonian
    \begin{equation}
    \begin{split}
        H
        &=
        m_\Omega\frac{\Omega(t)}{2}
        \sum_{j=1}^N
        e^{-{x_j^2}/{w^2}}\sigma_j^x\\
        &-\left[\delta(t)+\delta_{\rm qs}\right]
        \sum_{j=1}^N n_j^z
        +
        \sum_{j=1}^N\sum_{k<j}
        U_{jk}n_j^z n_k^z \, ,
    \end{split}
    \end{equation}
    and
    \begin{equation}
    \begin{split}
        L_\alpha
        &\in
        \left\{
            n_j^z,\sigma_j^-
            \,\middle|\,
            j=1,\dots,N
        \right\} \,,\\
        \gamma_\alpha &\in \{\Gamma,\gamma_\phi \ \forall \ j=1,\dots,N\} \, .
    \end{split}
    \end{equation}

    \item Apply the confusion matrix defined by $\FN$ and $\FP$ as in Eq.~\eqref{eq:Nconfmatrix} to the state probabilities extracted from the evolved density matrix $\rho(T)$.

    \item Average the resulting probabilities over the distribution of quasi-static detuning offsets, repeating points 2--3 for different values of $\delta_{\rm qs}$ (see App.~\ref{app:quadrature} for details on our implementation of this averaging procedure).

    \item Average the resulting probabilities over the positions of individual atoms or pairs of atoms, as appropriate for the dataset under consideration.
\end{enumerate}

\section{Simulation and results}\label{sec:results}
We now present the results obtained using the model introduced in Sec.~\ref{sec:model}. Since the noise parameters are \textit{a priori} unknown, we infer them through numerical optimization. Specifically, we compare the output of the simulation pipeline with the experimental data and optimize over the parameter set
\begin{equation}
    \left\{
        \FN,\FP,m_\Omega,w,\delta_0,\sigma_\delta,\Gamma,\gamma_\phi
    \right\} \, .
\end{equation}

Before proceeding, we note that some of the noise sources included in our model are expected to remain stable over relatively long timescales, whereas others may vary more rapidly.
We therefore divide the associated parameters into two categories.
The first category contains the parameters expected to be comparatively stable: the false-positive and false-negative detection probabilities $\FP$ and $\FN$, which are mostly determined by the detection procedure; the decay rate $\Gamma$, which is intrinsic to the atoms; and the laser waist $w$, which is set by the optical setup.
The second category contains the parameters expected to vary on shorter timescales: the dephasing rate $\gamma_\phi$, which generically stems from stochastic laser imperfections and environment coupling; the mean $\delta_0$ and standard deviation $\sigma_\delta$ of the quasi-static detuning distribution; and the amplitude multiplier $m_\Omega$.
Parameters in the first category are shared across multiple experimental datasets, whereas those in the second category are allowed to vary between datasets.

Here, an experimental dataset corresponds to a specific pulse sequence and register configuration. We recall that to reconstruct the dynamics at several evolution times, the same pulse sequence is executed with different truncation times, and, for each truncation time, the sequence is repeated over multiple shots to estimate the corresponding outcome probabilities. Thus, even parameters that are allowed to vary between datasets are assumed to remain approximately constant throughout the acquisition of a single dynamical dataset.

Overall, the optimization therefore involves a collection of parameters, some of which are shared among several datasets and others of which are dataset dependent.

Based on the dates and times at which the experimental datasets were acquired, we divide them into the following groups:
\begin{enumerate*}[label=(\arabic*)]
    \item \verb|dyn_0| and \verb|dyn_1|;
    \item \verb|dyn_2|, \verb|dyn_3|, and \verb|dyn_4|;
    \item \verb|dyn_5|;
    \item \verb|dyn_6|;
    \item \verb|dyn_7|;
    \item \verb|dyn_8|, \verb|dyn_9|, and \verb|dyn_10|.
\end{enumerate*}
Within each group, the long-timescale parameters are constrained to take common values, whereas the dataset-dependent parameters are optimized independently for each pulse sequence.

For quantifying the accuracy of our model for reproducing experimental data, we use as a metric the mean value of component-wise root mean square error; namely, for experimental and model probability vectors, respectively $\vec p_{\rm exp}$ and $\vec p_{\rm model}$, we compute
\begin{equation}
\begin{split}
    \oRMSE
    = \frac{1}{2^N D}\sum_{d=1}^{D}\sum_{k=1}^{2^N} \sqrt{\frac{1}{\lvert t_{\rm set}^{(d)}\rvert}\sum_{\tau\in t_{\rm set}^{(d)}}\left(p_{\rm exp}^{(d,\tau,k)}- p_{\rm model}^{(d,\tau,k)}\right)^2}
\end{split}
\end{equation}
where $p^{(d,\tau,k)}$ denotes the $k$-th component of the probability vector $\vec p$ obtained at the $\tau$-th observed time of the $d$-th experimental dataset, $t_{\rm set}^{(d)}$ denotes the set of times at which the $d$-th experimental dataset sequence has been observed, and $D$ amounts to the number of experimental datasets included in the computation.
This quantity has the direct interpretation of describing how much our model component-wise deviates on average from experimental data.
Moreover, the order of nested averages ensures that each experimental dataset has the same weight in the computation of $\oRMSE$, regardless of how many times $\tau$ have been observed.

A closely related quantity is obtained by replacing the innermost argument of $\oRMSE$ with the ratio between the observed deviation and its statistical error:
\begin{equation}
\begin{adjustbox}{max width=\linewidth}
    $\displaystyle
    \oRMSZ
    = \frac{1}{2^N D}\sum_{d=1}^{D}\sum_{k=1}^{2^N} \sqrt{\frac{1}{\lvert t_{\rm set}^{(d)}\rvert}\sum_{\tau\in t_{\rm set}^{(d)}}
    \frac{\left(p_{\rm exp}^{(d,\tau,k)}- p_{\rm model}^{(d,\tau,k)}\right)^2}{\varepsilon^2_{p_{\rm exp}^{(d,\tau,k)}}}} \, ,
    $
\end{adjustbox}
\end{equation}
where $\varepsilon_{p_{\rm exp}^{(d,\tau,k)}}$ denotes the statistical error (coming from experimental uncertainty) associated with $p_{\rm exp}^{(d,\tau,k)}$. The name $\oRMSZ$ reflects the fact that each term in the sum over $\tau$ is a squared Z-score. This quantity is closely related to the reduced chi-squared $\chi^2_\nu = \sum_j Z_j^2/\nu$, where $Z_j$ are the Z-scores of the data points and $\nu$ is the number of degrees of freedom, but it differs from it in two respects: it involves a square root, and it does not account for the number of free parameters of the model.
The $\oRMSZ$ can be interpreted as follows: a value below $1$ indicates that the model reproduces the data within the error bars on average, whereas a value above $1$ indicates deviations larger than the statistical uncertainty.

In the following, all of the error bars considered and displayed in plots have been obtained with a Bayesian approach to the estimation of statistical errors, whose details are reported in App.~\ref{app:bayes}.

Nonetheless, during the optimization we used the simpler squared-${\rm L}_2$ distance
\begin{equation}
    \sum_{d,\tau,k} \left( p_{\rm exp}^{(d,\tau,k)}- p_{\rm model}^{(d,\tau,k)}\right)^2
\end{equation}
as loss function, in order to avoid the average over nonlinear functions and maintain smoothness.

\begin{figure}
    \centering
    \includegraphics[width=0.95\linewidth]{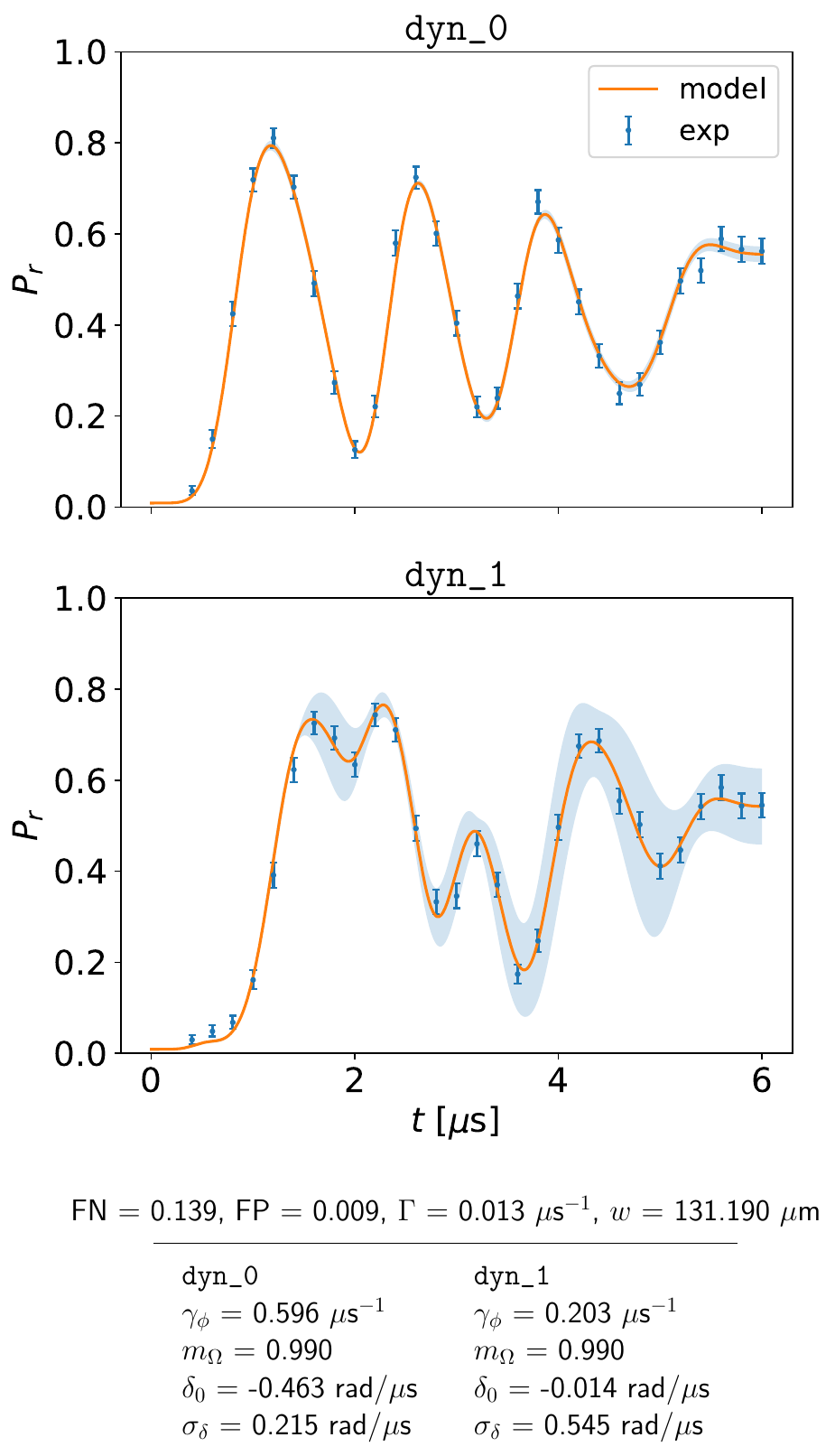}
    \cprotect\caption{Comparison between the experimental data obtained from datasets \verb|dyn_0| and \verb|dyn_1| (averaged over atoms' positions) and the predictions of our model with optimized parameters. Parameters listed above the horizontal line are shared between datasets, while those listed below it are fitted separately for each dataset. Error bars on the experimental data are obtained with the Bayesian estimation procedure described in App.~\ref{app:bayes}. The shaded area indicates the one-standard-deviation spread of the model prediction arising from the quasi-static detuning distribution with width $\sigma_\delta$.}
    \label{fig:1q-opt}
\end{figure}

\begin{figure*}
    \centering
    \includegraphics[width=0.95\linewidth]{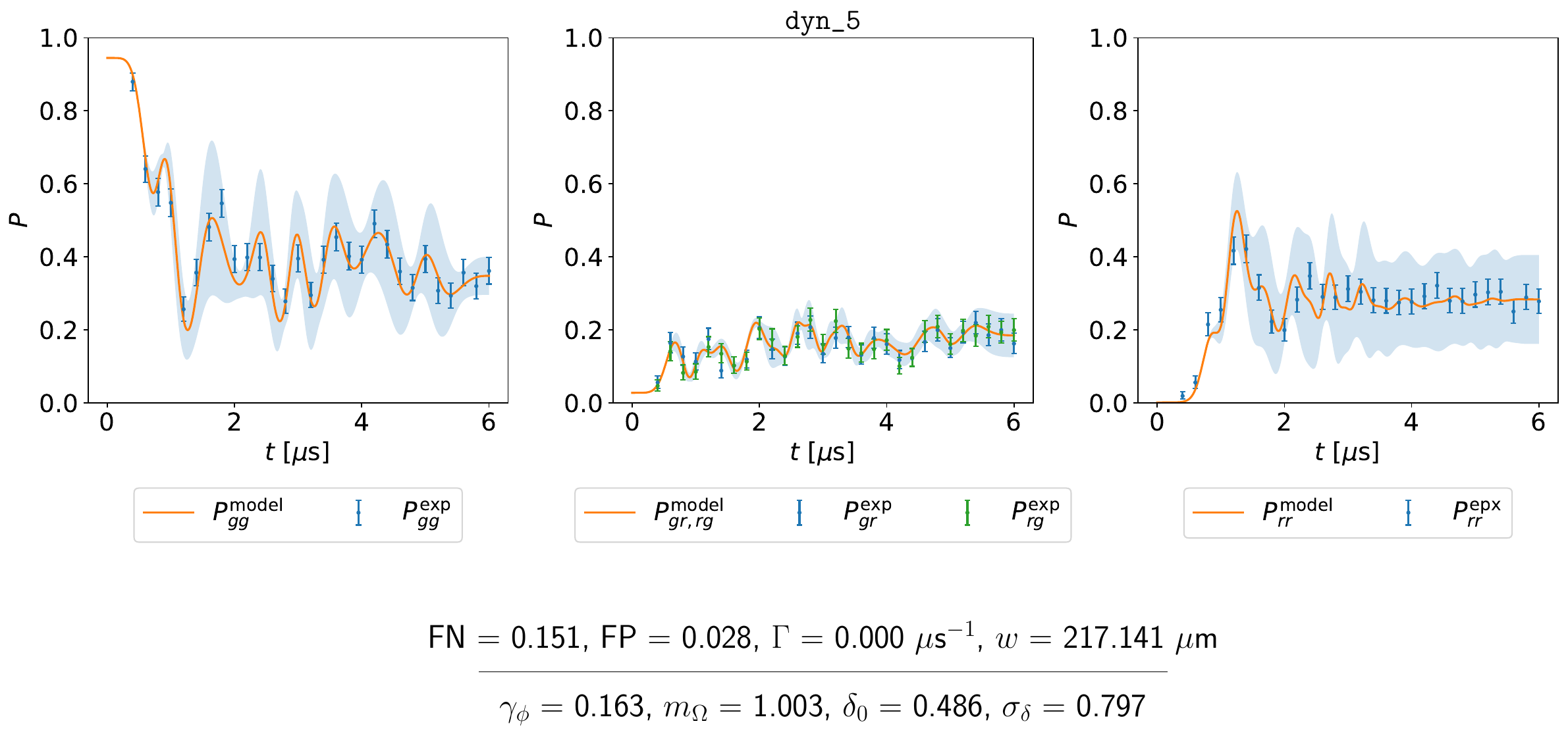}\\\vspace{2em}
    \includegraphics[width=0.95\linewidth]{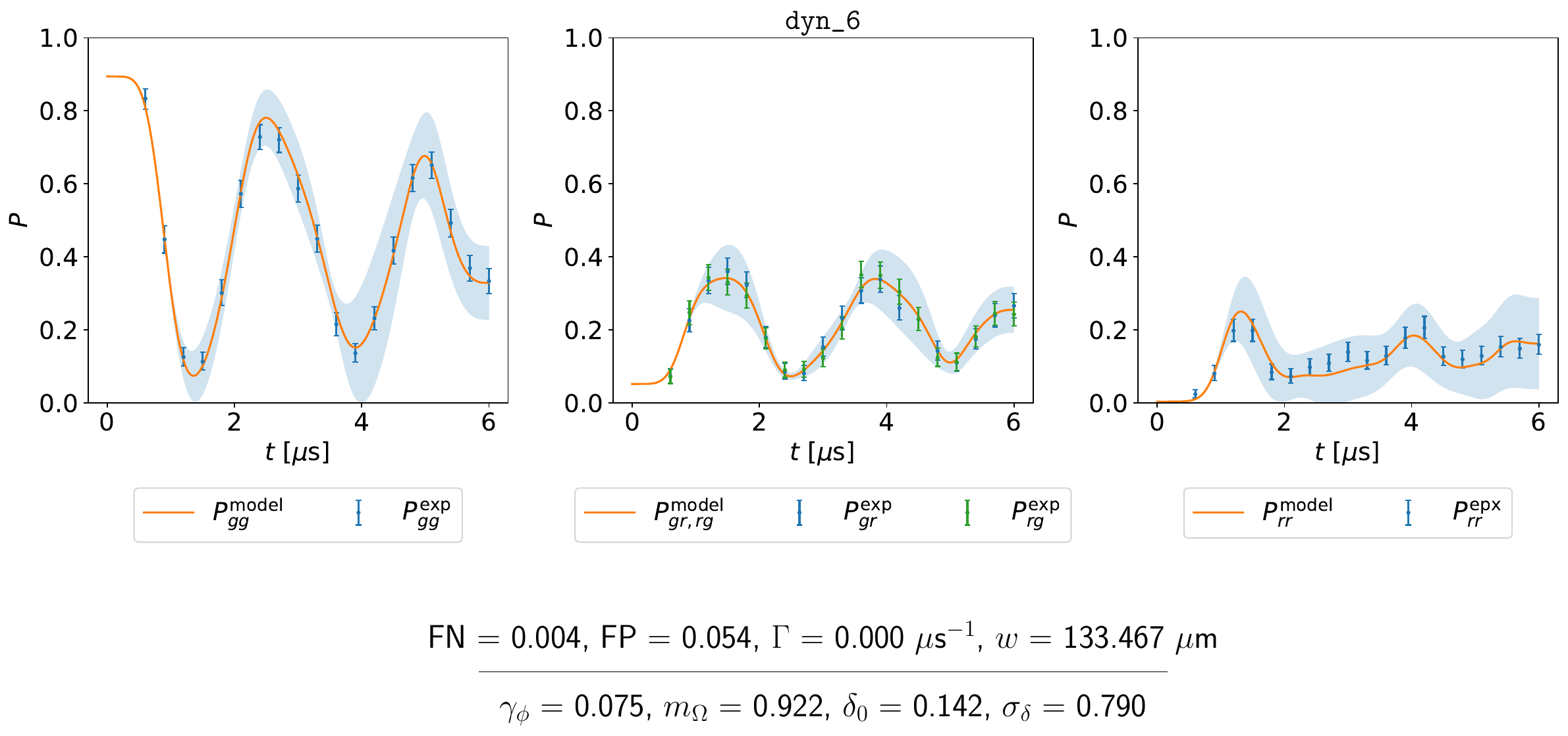}
    \cprotect\caption{Comparison between the experimental data obtained from datasets \verb|dyn_5| and \verb|dyn_6| (averaged over pairs of atoms) and the predictions of our model with optimized parameters. Each dataset has an independently fitted set of parameters. Error bars on the experimental data are obtained with the Bayesian estimation procedure described in App.~\ref{app:bayes}. The shaded area indicates the one-standard-deviation spread of the model prediction arising from the quasi-static detuning distribution with width $\sigma_\delta$.}
    \label{fig:2q-opt}
\end{figure*}

Fig.~\ref{fig:1q-opt} compares the optimized dynamics predicted by our model with the experimental data for a representative single-qubit dataset. Similarly, Fig.~\ref{fig:2q-opt} presents the optimized dynamics for representative two-qubit datasets. Extended data and figures can be found in App.~\ref{app:extended-data}. In both cases, the model predictions are obtained using the noise parameters inferred through the optimization procedure described above.
As results display, with our model we have been able to reproduce the system's dynamics accurately, obtaining a value of $\oRMSZ<1$ for every case investigated; that is, the dynamics implemented by our model always reproduces, on average, the experimental one within the error bars.

For the single-qubit datasets, Fig.~\ref{fig:1q-opt} shows the results of the joint analysis of experiments \verb|dyn_0| and \verb|dyn_1|, which our model describes with $\oRMSE\approx0.078$ and $\oRMSZ\approx0.692$.
Furthermore, we highlight that the optimal noise parameters are physically realistic and consistent with expectations for this type of hardware: both measurement errors are limited to the order of a few percent; the decay rate is very small; the laser waist is about $2.5$ times the maximum radius $R_{\rm max}$ of the register; the dephasing rates correspond to timescales of a few microseconds; the amplitude multipliers are close to $1$; the detuning offsets are limited; and the standard deviations of the quasi-static detunings are finite but small.
Notably, $m_\Omega$, although free to vary independently for each experimental dataset, converge to close values in both, and similarly also $\delta_0$ is smaller than $1$ for each experiment. The exceptions are $\gamma_\phi$ and $\sigma_\delta$. As discussed above, however, these two parameters have very similar empirical effects on the dynamics and are therefore strongly correlated: \verb|dyn_0| exhibits a larger $\gamma_\phi$ and a smaller $\sigma_\delta$, while the opposite holds for \verb|dyn_1|.

\begin{figure}
    \centering
    \includegraphics[width=0.95\linewidth]{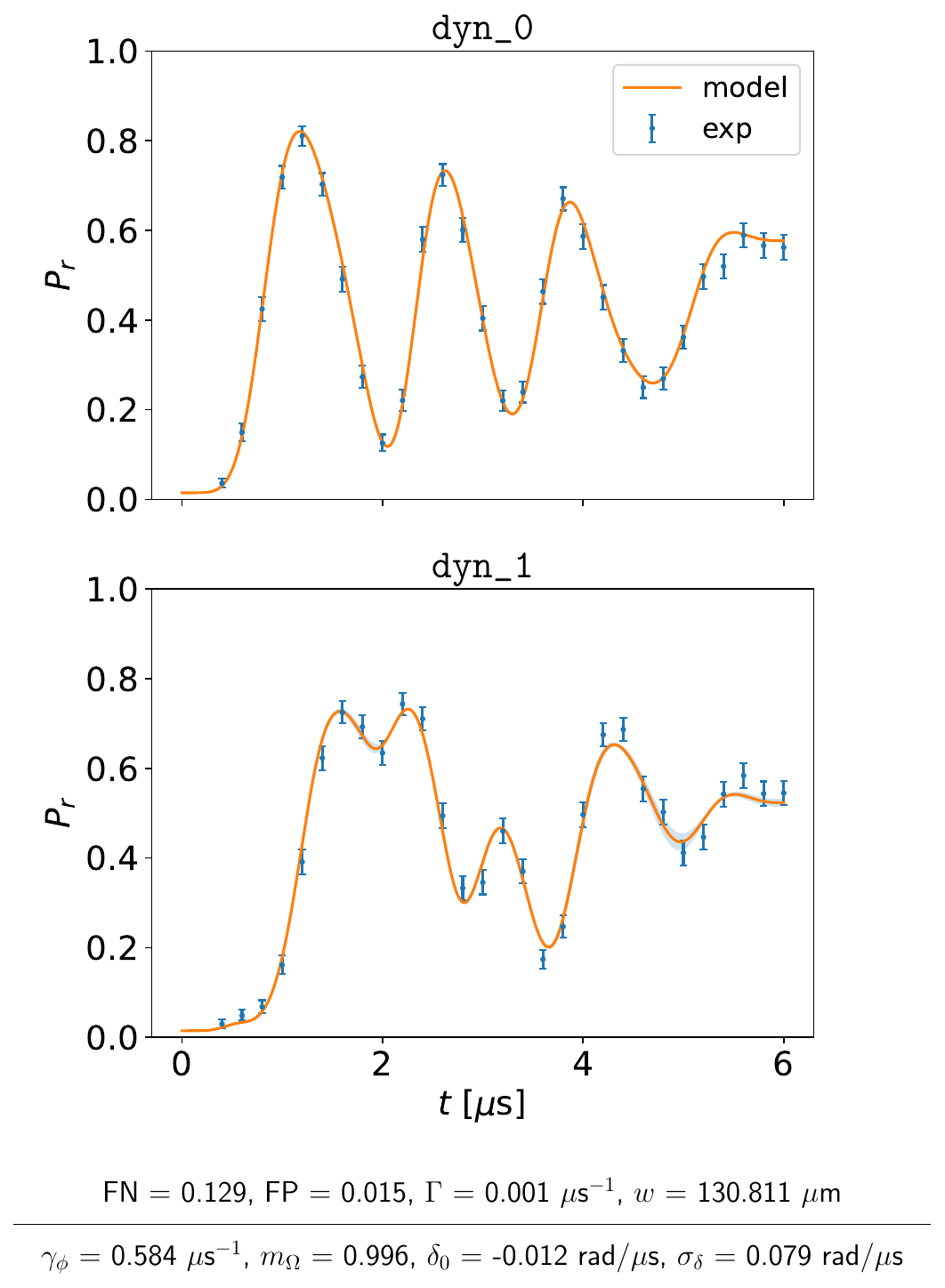}
    \cprotect\caption{Comparison between the experimental data obtained from datasets \verb|dyn_0| and \verb|dyn_1| (averaged over atoms' positions) and the predictions of our model with optimized parameters. All parameters are shared between datasets. Error bars on the experimental data are obtained with the Bayesian estimation procedure described in App.~\ref{app:bayes}. The shaded area indicates the one-standard-deviation spread of the model prediction arising from the quasi-static detuning distribution with width $\sigma_\delta$.}
    \label{fig:1q-opt-common}
\end{figure}

\begin{figure}
    \centering
    \includegraphics[width=0.95\linewidth]{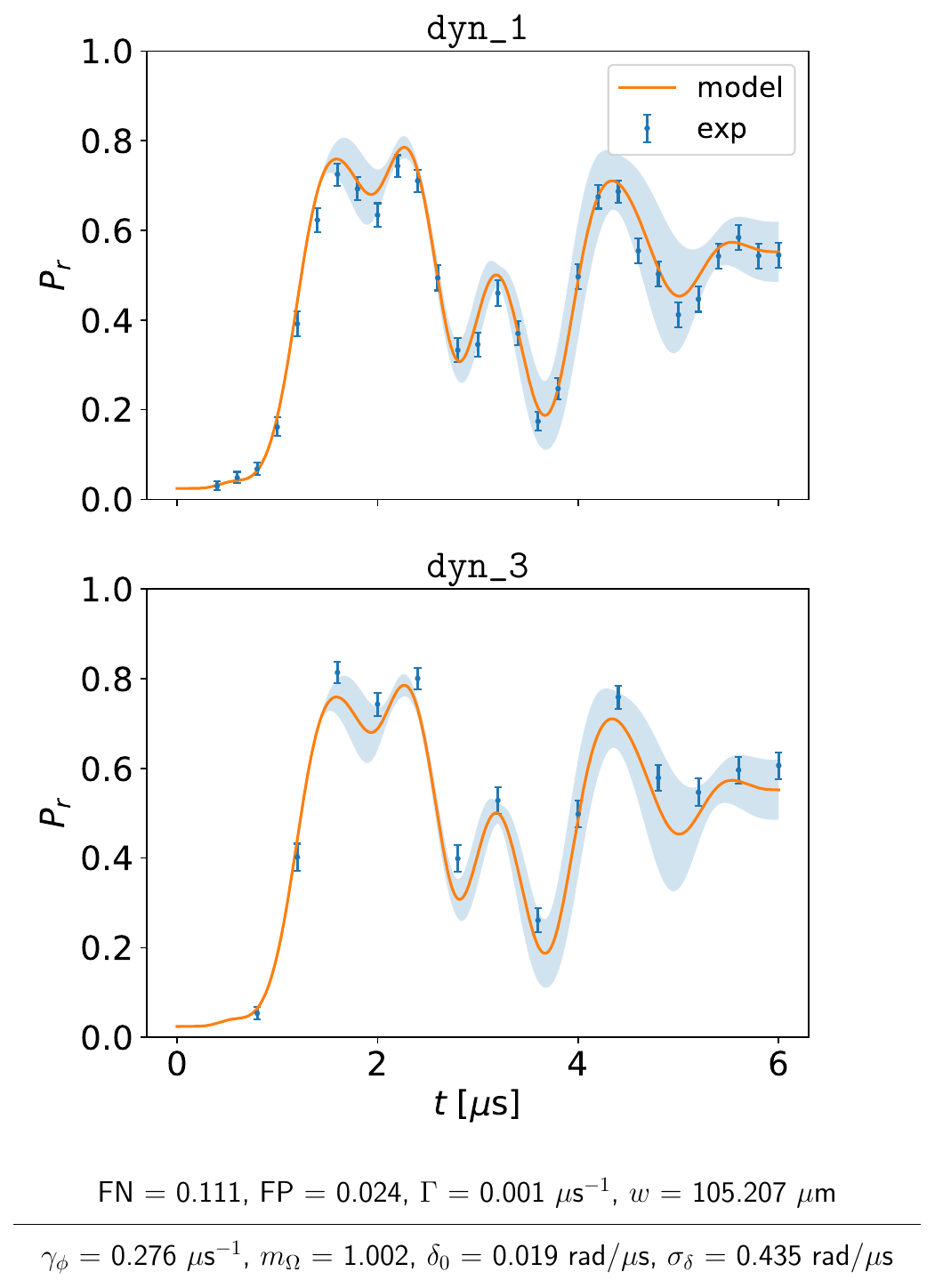}
    \cprotect\caption{Comparison between the experimental data obtained from datasets \verb|dyn_1| and \verb|dyn_3| (averaged over atoms' positions) and the predictions of our model with optimized parameters. All parameters are shared between datasets. Error bars on the experimental data are obtained with the Bayesian estimation procedure described in App.~\ref{app:bayes}. The shaded area indicates the one-standard-deviation spread of the model prediction arising from the quasi-static detuning distribution with width $\sigma_\delta$.}
    \label{fig:1q-common-diff}
\end{figure}

Given these considerations, and since \verb|dyn_0| and \verb|dyn_1| were performed consecutively on Fresnel, i.e.~within a short time window, we also fitted both datasets with all noise parameters shared. The results, reported in Fig.~\ref{fig:1q-opt-common}, show that the dynamics is still well reproduced, now with $\oRMSE\approx0.080$, which is just slightly higher than in the previous case.
Compared with Fig.~\ref{fig:1q-opt}, $\gamma_\phi$ and $\sigma_\delta$ converge to common values, while the other parameters remain comparable.
This confirms that, at least at a phenomenological level and at the available resolution, some parameters, such as $\gamma_\phi$ and $\sigma_\delta$, are strongly correlated for this type of experiment, and partially compensate for each other in the fit. It also indicates that a single set of noise parameters suffices to describe experiments performed close in time, i.e.~that noise processes remain stable for the time needed to acquire several experimental datasets.
The same approach is less successful for experiments carried out far apart in time, suggesting that there is indeed a drift in the strength of noise sources over longer periods. As an example, Fig.~\ref{fig:1q-common-diff} shows a shared-parameter fit of \verb|dyn_1| and \verb|dyn_3|, two datasets that even use the same pulse sequence. The resulting dynamics deviates qualitatively more from the experimental data and is more strongly biased (in particular, the model dynamics for \verb|dyn_0| is biased upwards with respect to the experimental data, while the one for \verb|dyn_3| downwards), although it remains acceptable in terms of $\oRMSZ$ and $\oRMSE$.

Similarly, for the two-qubit datasets, Fig.~\ref{fig:2q-opt} shows the results for experiments \verb|dyn_5| and \verb|dyn_6|, which in this case have been analyzed separately since they were performed on different days, yielding $\oRMSZ\approx0.552,0.536$ and $\oRMSE\approx0.058,0.054$, respectively.
As in the single-qubit case, the optimized parameter values are realistic, and the fits are good both qualitatively and quantitatively.
One clear difference from the single-qubit case, however, is the larger standard deviation of the populations induced by a greater $\sigma_\delta$, shown as a shaded area in the plots. Although this broadens the predicted spread of the populations, a finite $\sigma_\delta$ is in some cases necessary, because it affects not only the variance of the results but also their mean over the quasi-static detuning distribution.
This suggests that two-qubit experiments are more sensitive to $\sigma_\delta$ than single-qubit ones, making its effect both stronger and easier to detect.
In support of this, for both \verb|dyn_5| and \verb|dyn_6| we were unable to reproduce the experimental dynamics without a quasi-static detuning distribution, whereas including it yielded very accurate fits; this is particularly valid for the fast oscillations of \verb|dyn_5|, which also exhibit a strong loss of contrast in $P_{rr}$.

\section{Summary and conclusions}\label{sec:conclusions}

In summary, in this paper we have studied the characterization of cloud-accessible quantum hardware. Taking advantage of the growing availability of such platforms, we used Fresnel, a neutral-atom quantum simulator provided by Pasqal.
As remote users, we could not run calibration or diagnostic routines directly on the hardware, instead we could only execute experiments and read out the measurement outcomes in the computational basis returned by the machine.
To exploit this input--output data, we needed to design a protocol capable of revealing information about the relevant imperfections affecting the device.

To this end, we first reviewed the noise sources expected to be most influential for this type of hardware, focusing on their effects on the quantum dynamics and on the signatures each one leaves.
We then designed experiments able to reveal these signatures and their strength: to this purpose, time-resolved experiments proved particularly useful, as they exploit the information encoded in the dynamical behavior of the system.
Hence, we examined the experimental outcomes under the light of noise signatures considered.

Having identified the main noise signatures, we used this physical information to build an effective model that extends the ideal Rydberg-Hamiltonian dynamics. In particular, we included measurement errors; perturbations of the drives, in the form of an amplitude multiplier and a detuning offset; spatial inhomogeneity of the laser, through a finite waist; shot-to-shot fluctuations, in the form of quasi-static detuning noise; spontaneous emission from $\ket{r}$ to $\ket{g}$; and dephasing.
Since the strength of each of these effects is unknown \textit{a priori}, we used numerical optimization to tune corresponding noise parameters in order to fit the experimental data.
This procedure closely reproduced the experimental data, with deviations that remained, on average, within the experimental error bars.
Moreover, the optimized noise parameters took physically reasonable values, consistent across the different datasets, which supports the reliability of our approach.

To conclude, we have shown how physical intuition about typical noise sources and their effects, combined with time-resolved data and numerical optimization, can guide the construction of an effective model of the hardware's behavior.
Although we focused here on a neutral-atom quantum simulator, the approach is general and, once tailored to the features of the specific platform, can be readily adapted to other types of quantum machines.

\begin{acknowledgments}
We acknowledge funding under Horizon Europe programme HORIZON-CL4-2022-QUANTUM-02-SGA via the Grant No. 101113690 (PASQuanS2.1). A.M. acknowledges CINECA for having provided access to both quantum and classical computational resources.
\end{acknowledgments}

\appendix

\section{Single-qubit dynamics}
In this Appendix, we report the detailed calculations of the dynamics in the presence of various noise sources. For simplicity, we focus on single-qubit dynamics under constant drives, which lead to analytically tractable equations. The generalization to more general settings is usually straightforward.

\subsection{Resonant drive with dephasing}\label{app:dephasing}
We analyze here the dynamics of a single atom resonantly driven in presence of dephasing.
This is described by a Lindblad master equation with the jump operator $L_1=L=\sigma^z$ and its associated dephasing rate $\gamma_\phi$:
\begin{equation}
    \dot \rho = -i\frac{\Omega}{2}[\sigma^x,\rho] + {\gamma_\phi}(\sigma^z\rho\sigma^z-\rho) \, ,
\end{equation}
where the subscript enumerating the (only) atom is omitted.
Rewriting $\rho=(\mathbbm{1}+s_x\sigma^x+s_y\sigma^y+s_z\sigma^z)/2$, with $s_{\alpha}=\langle\sigma^\alpha\rangle$ ($\alpha=x,y,z$) indicating the expectation values, the dynamics then becomes
\begin{equation}
    \begin{cases}
        \dot s_x= -2\gamma_\phi s_x\\
        \dot s_y = -\Omega s_z-2\gamma_\phi s_y\\
        \dot s_z = \Omega s_y
    \end{cases} \, ,
\end{equation}
where we have used the Pauli matrices' relations $[\sigma^\alpha,\sigma^\beta]=2i\epsilon_{\alpha\beta\gamma}\sigma^\gamma$ and $\sigma^z\sigma^{x,y}\sigma^z=-\sigma^{x,y}$.
Differentiating the last equation with respect to time and expliciting in $s_z$, we obtain:
\begin{equation}
    \ddot s_z + 2\gamma_\phi \dot s_z + \Omega^2 s_z = 0
\end{equation}
which is a damped harmonic oscillator. Considering the underdamped case $\gamma_\phi<\Omega$ with initial conditions $s_z(0)=1$ (i.e.~$\ket{\psi(0)}=\ket g$ and $\dot s_z(0)=0$ (coming from $\dot s_z = \Omega s_y|_{t=0}$ with $\ket{\psi(0)}=\ket g$), the solution is
\begin{equation}
\begin{split}
    s_z(t) = e^{-\gamma_\phi t} \Bigg[& \cos\left(\sqrt{\Omega^2-\gamma_\phi^2}\;t\right) + \\
    &+ \frac{\gamma_\phi}{\sqrt{\Omega^2-\gamma_\phi^2}}\sin\left(\sqrt{\Omega^2-\gamma_\phi^2}\;t\right) \Bigg] \, ,
    \label{eq:resonant-dephasing}
\end{split}
\end{equation}
which can be further simplified into
\begin{equation}
    s_z(t) = e^{-\gamma_\phi t} \sqrt{1+a^2} \cos\left(\sqrt{\Omega^2-\gamma_\phi^2}\;t - \varphi \right)
    \label{eq:resonant-dephasing-2}
\end{equation}
where $a={\gamma_\phi}/{\sqrt{\Omega^2-\gamma_\phi^2}}$ and $\varphi=\arctan a$.
In case the jump operator is given by $L=n^z=(\mathbbm{1}-\sigma^z)/2$, the same result hold upon substitution $\gamma_\phi\mapsto\gamma_\phi/4$.

We highlight that Eq.~\eqref{eq:resonant-dephasing-2} can be used to compute the phase shift between two signals of different frequency. In particular, the presence of dephasing lowers the frequency of Rabi oscillations in such a way that the phase accumulated over time with respect to unitary dynamics is
\begin{equation}
    \Delta\varphi=\left(\sqrt{\Omega^2-\gamma_\phi^2}-\Omega\right)t + \varphi \, .
    \label{eq:phase-dephasing}
\end{equation}

\subsection{Resonant drive with decay}\label{app:decay}

We analyze here the dynamics of a single atom undergoing spontaneous decay from the Rydberg state to the ground state, described by the jump operator $L_1=L=\sigma^-=\ketbra{g}{r}$ ($L^\dag=\sigma^+=\ketbra{r}{g}$), with its associated decay rate $\Gamma$. The Lindblad master equation reads
\begin{equation}
    \dot \rho = -i\frac{\Omega}{2}[\sigma^x,\rho] + \Gamma \left(\sigma^-\rho\sigma^+-\frac12\{\sigma^+\sigma^-,\rho\}\right) \, ,
\end{equation}
where we again omit the subscript enumerating the atom.
Rewriting $\rho=(\mathbbm{1}+s_x\sigma^x+s_y\sigma^y+s_z\sigma^z)/2$, the dynamics becomes
\begin{equation}
    \begin{cases}
        \dot s_x &= -\Gamma s_x/2\\
        \dot s_y &= -\Omega s_z -\Gamma s_y/2\\
        \dot s_z &= \Omega s_y +\Gamma(1-s_z)
    \end{cases} \, ,
\end{equation}
where we have used that $L^\dag L = \sigma^+\sigma^- = \ketbra{r}{r} = (\mathbbm{1}-\sigma^z)/2$, $\sigma^-\sigma^{x,y}\sigma^+=0$, $\sigma^-\sigma^z\sigma^+=-\ketbra{g}{g}=-(\mathbbm{1}+\sigma^z)/2$.
Contrary to the dephasing case, the equation for $s_z$ now contains an inhomogeneous term, reflecting the fact that decay repopulates the ground state.
Differentiating the last equation with respect to time and eliminating $s_y$, we obtain
\begin{equation}
    \ddot s_z + \frac{3\Gamma}{2} \dot s_z + \left(\Omega^2+\frac{\Gamma^2}{2}\right) s_z = \frac{\Gamma^2}{2} \, ,
\end{equation}
which is a damped harmonic oscillator subject to a constant force.
Considering the underdamped case $\Gamma<4\Omega$ with initial conditions $s_z(0)=1$ (i.e.~$\ket{\psi(0)}=\ket g$) and $\dot s_z(0)=0$ (following from $s_y(0)=0$ and $s_z(0)=1$), the solution is
\begin{equation}
\begin{split}
    s_z(t) &= \frac{\Gamma^2}{\Gamma^2+2\Omega^2} \\
    &+ \frac{2\Omega^2}{\Gamma^2+2\Omega^2}\, e^{-\frac{3\Gamma}{4} t} \Bigg[\cos\left(\Omega_\Gamma t\right)
    + \frac{3\Gamma}{4\Omega_\Gamma}\sin\left(\Omega_\Gamma t\right) \Bigg] \, ,
    \label{eq:resonant-decay}
\end{split}
\end{equation}
where $\Omega_\Gamma=\sqrt{\Omega^2-\Gamma^2/16}$. This can be further simplified into
\begin{equation}
\begin{split}
    s_z(t) &= \frac{\Gamma^2}{\Gamma^2+2\Omega^2} \\
    &+ \frac{2\Omega^2}{\Gamma^2+2\Omega^2}\, e^{-\frac{3\Gamma}{4} t} \sqrt{1+b^2} \cos\left(\Omega_\Gamma t - \varphi \right) \, ,
    \label{eq:resonant-decay-2}
\end{split}
\end{equation}
where $b={3\Gamma}/{(4\Omega_\Gamma)}$ and $\varphi=\arctan b$.
Unlike pure dephasing, which drives the atom towards the maximally mixed state $s_z\to0$, spontaneous decay drives it towards a steady state with finite ground-state bias $s_z(\Gamma t\gg1)>0$, while the oscillations are damped at rate $3\Gamma/4$.

\subsection{Quasi-static detuning noise}\label{app:quasi-static}
We report in this Section detailed computations about the dynamics of a single atom in presence of quasi-static detuning noise.
In particular, we consider random values of detuning sampled from the Gaussian distribution with mean $\delta_0$ and standard deviation $\sigma_\delta$
\begin{equation}
    \delta_{\rm qs}\sim \mathcal{N}(\delta_0,\sigma_\delta^2) \, ,
\end{equation}
which modify the nominal detuning $\delta(t)$ as
\begin{equation}
\delta(t) \longmapsto \delta(t)+\delta_{\rm qs} \, .
\end{equation}
Although each single evolution is still unitary, since $\delta_{\rm qs}$ only enters as a perturbation of the driving field, experimental observations are obtained by averaging over several system evolutions, each one with a different value of $\delta_{\rm qs}$; hence, the effective ensemble-averaged dynamics is generically non-unitary.

In formulae, let $P_r(t\mid \delta_{\rm qs})$ denote the probability of detecting an atom in $\ket{r}$ for a given value of $\delta_{\rm qs}$, assuming the atom is initially prepared in $\ket{g}$. If we restrict ourselves to the case of constant drives (i.e.~time-independent $\Omega$ and $\delta$), we have from Eq.~\eqref{eq:Rabi-oscillations-detuned} that
\begin{equation}
\begin{split}
    P_r(t\mid \delta_{\rm qs})
    &= \frac{\Omega^2}{\Omega^2+\widetilde\delta^2}
    \sin^2\left(
        \frac{t}{2}\sqrt{\Omega^2+\widetilde\delta^2}
    \right) \\
    &= \frac12\frac{\Omega^2}{\Omega^2+\widetilde\delta^2}
    \left[
    1-\cos\left(
        {t}\sqrt{\Omega^2+\widetilde\delta^2}
    \right) \right]\, ,
\end{split}
\end{equation}
where we denoted $\widetilde\delta=\delta+\delta_{\rm qs}$.
The ensemble-averaged value is consequently
\begin{equation}
\begin{split}
    P_r^{\rm avg}(t)
    &= \int_{-\infty}^\infty p(\delta_{\rm qs}) \: P_r(t\mid\delta_{\rm qs}) \: {\rm d}\delta_{\rm qs} \\
    &= \int_{-\infty}^\infty \frac{e^{-({\delta_{\rm qs}-\delta_0)^2}/{(2\sigma_\delta^2)}}}{\sqrt{2\pi\sigma_\delta^2}} \: P_r(t\mid\delta_{\rm qs}) \: {\rm d}\delta_{\rm qs}\\
    &= \Big\langle P_r(t\mid\delta_{\rm qs})\Big\rangle \, ,
\end{split}
\end{equation}
where $\langle\cdot\rangle$ here denotes the expectation value with respect to the distribution of $\delta_{\rm qs}$.
In order to proceed with the explicit computation of $P_r^{\rm avg}(t)$, we need to address separately three different cases, namely the resonant one (i.e.~with $\delta=\delta_0=0$), the detuned one (i.e.~with $\delta,\delta_0\neq0$), and the intermediate one bridging between the two, since the noise affects the dynamics differently: on resonance it enters the generalized Rabi frequency only at second order in $\delta_{\rm qs}$, while off resonance it already enters at first order.
We remark that the detuned case can be modeled absorbing the offset into the drive, namely considering a nominal detuning drive $\delta\mapsto\delta+\delta_0$ perturbed by $\delta_{\rm qs}\sim{\cal N}(0,\sigma_\delta^2)$.

\subsubsection{Resonant case}
Let us analyze here the case on resonance, namely considering $\delta=0$ and $\delta_0=0$.
In this case we have
\begin{equation}
    P_r(t\mid \delta_{\rm qs})
    = \frac12\frac{\Omega^2}{\Omega^2+\delta_{\rm qs}^2}
    \left[
    1-\cos\left(
        {t}\sqrt{\Omega^2+\delta_{\rm qs}^2}
    \right) \right]\, ,
\end{equation}
and
\begin{equation}
\begin{split}
    P_r^{\rm avg}(t)
    &= \int_{-\infty}^\infty \frac{e^{-{\delta_{\rm qs}^2}/{(2\sigma_\delta^2)}}}{\sqrt{2\pi\sigma_\delta^2}} \: P_r(t\mid\delta_{\rm qs}) \: {\rm d}\delta_{\rm qs}\\
    &= \Big\langle P_r(t\mid\delta_{\rm qs})\Big\rangle \, .
\end{split}
\end{equation}
In the weak-disorder regime $\sigma_\delta\ll\Omega$ we can approximate for small values of $\delta_{\rm qs}$
\begin{equation}
\begin{split}
    \sqrt{\Omega^2+\delta_{\rm qs}^2}&\approx\Omega+\frac{\delta_{\rm qs}^2}{2\Omega} \, , \\
    \frac{\Omega^2}{\Omega^2+\delta_{\rm qs}^2}&\approx 1-\frac{\delta_{\rm qs}^2}{\Omega^2} \, ,
\end{split}
\end{equation}
thus getting
\begin{equation}
\begin{split}
    P_r(t\mid \delta_{\rm qs})
    &= \frac12\frac{\Omega^2}{\Omega^2+\delta_{\rm qs}^2}
    \left[
    1-\cos\left(
        {t}\sqrt{\Omega^2+\delta_{\rm qs}^2}
    \right) \right] \\
    &= \frac12\frac{\Omega^2}{\Omega^2+\delta_{\rm qs}^2}
    \left[
    1-\Re\left\{\exp\left(
        i{t}\sqrt{\Omega^2+\delta_{\rm qs}^2}
    \right)\right\} \right] \\
    &\approx \frac12
    \left[
    1-\Re\left\{e^{i\Omega t} e^{i\delta_{\rm qs}^2t/(2\Omega)}\right\} \right] \, ,
\end{split}
\end{equation}
$\Re\{x\}$ denoting the real part of $x$, and its corresponding ensemble-averaged value
\begin{equation}
    P_r^{\rm avg}(t) \approx
    \frac12
    \left[
    1-\Re\left\{e^{i\Omega t} \left\langle e^{i\delta_{\rm qs}^2t/(2\Omega)}\right\rangle\right\} \right] \, .
\end{equation}
This last expression can be further manipulated recalling that for a zero-mean Gaussian distribution it holds
\begin{equation}
    \langle e^{ia\delta_{\rm qs}^2}\rangle = (1-2ia\sigma_\delta^2)^{-1/2} \, ,
\end{equation}
and using the identity
\begin{equation}
    (1-ix)^{-1/2} = (1+x^2)^{-1/4}\exp\left(\frac i2\arctan x\right) \, ,
\end{equation}
thus getting
\begin{equation}
    P_r^{\rm avg}(t) \approx \frac12 \left[\vphantom{\frac12}1- C(t)\cos\big(\Omega t + \varphi(t)\big)\right] \, ,
\end{equation}
with
\begin{equation}
\begin{split}
    C(t) &= \left[1+\left(\frac{\sigma_\delta^2t}{\Omega}\right)^2\right]^{-1/4}\, , \\
    \varphi(t) &= \frac12\arctan\left(\frac{\sigma_\delta^2t}{\Omega}\right) \, .
    \label{eq:quasi-static-resonant-envelope-phase}
\end{split}
\end{equation}
We obtain in this regime that, under the assumption of weak disorder, we have a loss of contrast with algebraic envelope $C(t)\sim t^{-1/2}$ for $t\gg\Omega/\sigma_\delta^2$.
At late times the population approaches
\begin{equation}
    P_r^{\rm avg}(t\gg\Omega/\sigma_\delta^2) = \frac12 \, ,
\end{equation}
up to corrections of order $\sigma_\delta^2/\Omega^2$ neglected in the approximation of the amplitude prefactor.

\subsubsection{Detuned case}
Consider now the detuned case, namely with $\delta\neq0$. Without loss of generality, we can assume $\delta_0=0$ since we can include its contribution inside $\delta$ itself.
In this case we have
\begin{equation}
\begin{split}
    P_r(t\mid \delta_{\rm qs})
    = \frac12&\frac{\Omega^2}{\Omega^2+(\delta+\delta_{\rm qs})^2} \times \\
    &\qquad\left[
    1-\cos\left(
        {t}\sqrt{\Omega^2+(\delta+\delta_{\rm qs})^2}
    \right) \right]\, ,
\end{split}
\end{equation}
and
\begin{equation}
    P_r^{\rm avg}(t)
    = \Big\langle P_r(t\mid\delta_{\rm qs})\Big\rangle \, ,
\end{equation}
with the same distribution for $\delta_{\rm qs}$ as in the resonant case.
In the weak-disorder regime $\sigma_\delta\ll\Omega$, and further assuming $\sigma_\delta\ll\lvert\delta\rvert$, we can approximate for small values of $\delta_{\rm qs}$
\begin{equation}
\begin{split}
    \sqrt{\Omega^2+(\delta+\delta_{\rm qs})^2}&\approx\widetilde\Omega+\frac{\delta}{\widetilde\Omega}\delta_{\rm qs} \, , \\
    \frac{\Omega^2}{\Omega^2+(\delta+\delta_{\rm qs})^2}&\approx \frac{\Omega^2}{\widetilde\Omega^2} \, ,
\end{split}
\end{equation}
where we denoted $\widetilde\Omega=\sqrt{\Omega^2+\delta^2}$.
Note that the additional condition $\sigma_\delta\ll\lvert\delta\rvert$ is essential for motivating the expansion above; in the case of $\sigma_\delta$ comparable to $\lvert\delta\rvert$, indeed, the expansion should be performed for small values of $\delta+\delta_{\rm qs}$ (with respect to $\Omega$), leading to a different result (see App.~\ref{app:quasi-stati-near-resonant} for this case). Because of this, the limit $\delta\to0$ of the results below does not reproduce the resonant case.
The ensemble-average Rydberg population is then
\begin{equation}
    P_r^{\rm avg}(t) \approx
    \frac{\Omega^2}{2\widetilde\Omega^2}
    \left[
    1-\Re\left\{e^{i\widetilde\Omega t} \left\langle e^{i(\delta/\widetilde\Omega)\delta_{\rm qs}t}\right\rangle\right\} \right] \, .
\end{equation}
Using that
\begin{equation}
    \langle e^{ia\delta_{\rm qs}}\rangle = \exp\left[-\frac12a^2\sigma_\delta^2\right] \, ,
\end{equation}
we get
\begin{equation}
    P_r^{\rm avg}(t) \approx
    \frac12\frac{\Omega^2}{\Omega^2+\delta^2}
    \left[
    1-C(t)\cos\left(t\sqrt{\Omega^2+\delta^2}\right) \right] \, ,
    \label{eq:qs-pr-detuned}
\end{equation}
with
\begin{equation}
    C(t) = \exp\left[-\frac12\left(\frac{\delta\sigma_\delta t}{\sqrt{\Omega^2+\delta^2}}\right)^2\right] \, .
    \label{eq:qs-envelope-detuned}
\end{equation}
In contrast with the resonant case, we see here that we have a Gaussian envelope damping the oscillations, $C(t)=e^{-(t/T_2^*)^2}$ with dephasing time $T_2^*=\sqrt2\,\widetilde\Omega/(\lvert\delta\rvert\sigma_\delta)$ (note that this is a different dephasing mechanism from the Markovian one described in Sec.~\ref{sec:dephasing} and App.~\ref{app:dephasing}, which produces an exponential envelope instead). The late-time dynamics approaches the population
\begin{equation}
    P_r^{\rm avg}(t\gg T_2^*) = \frac12\frac{\Omega^2}{\Omega^2+\delta^2} \, ,
\end{equation}
which is strictly smaller than $1/2$ for $\delta\neq0$ due to the presence of the detuning clipping the excitation amplitude.

\subsubsection{Near-resonant case}\label{app:quasi-stati-near-resonant}
We finally consider the case in which we have non-zero systematic detuning offset but the condition $\sigma_\delta\ll|\delta|$ is not met, i.e.~for $\sigma_\delta\ll\Omega$ and $\sigma_\delta$ comparable to $\delta$, which implies $|\delta|\ll\Omega$ as well. This can happen, for example, when the quasi-static detuning noise is sampled from
\begin{equation}
    \delta_{\rm qs} \sim {\cal N}(\delta_0,\sigma_\delta^2)
\end{equation}
with small $\lvert\delta_0\rvert>0$, and the detuning drive $\delta$ is set to $0$.
The expansion for small values of $\delta_{\rm qs}$ is the same as in the resonant case
\begin{equation}
\begin{split}
    \sqrt{\Omega^2+\delta_{\rm qs}^2}&\approx\Omega+\frac{\delta_{\rm qs}^2}{2\Omega} \, , \\
    \frac{\Omega^2}{\Omega^2+\delta_{\rm qs}^2}&\approx 1-\frac{\delta_{\rm qs}^2}{\Omega^2} \, ,
\end{split}
\end{equation}
and so the ensemble-averaged population
\begin{equation}
    P_r^{\rm avg}(t) \approx
    \frac12
    \left[
    1-\Re\left\{e^{i\Omega t} \left\langle e^{i\delta_{\rm qs}^2t/(2\Omega)}\right\rangle\right\} \right] \, .
\end{equation}
The difference comes in the evaluation of the average, which is now over a Gaussian distribution with non-zero mean. In this case it holds
\begin{equation}
    \left\langle e^{ib u^2}\right\rangle
    = \frac{1}{\sqrt{1-2ib\sigma_\delta^2}}
    \exp\left[\frac{ib\delta^2}{1-2ib\sigma_\delta^2}\right] \, ,
\end{equation}
which leads to the result
\begin{equation}
    P_r^{\rm avg}(t) \approx \frac12 \left[\vphantom{\frac12}1- C(t)\cos\big(\Omega t + \varphi(t)\big)\right] \, ,
\end{equation}
with
\begin{equation}
\begin{split}
    C(t) &= \left(1+\zeta^2\right)^{-1/4}
    \exp\left[-\frac{\delta^2\sigma_\delta^2t^2}{2\Omega^2\left(1+\zeta^2\right)}\right] \, , \\
    \varphi(t) &= \frac12\arctan\zeta + \frac{\delta^2t}{2\Omega\left(1+\zeta^2\right)} \, ,
\end{split}
\end{equation}
where $\zeta=\sigma_\delta^2t/\Omega$.
This expression bridges the two previous cases.
For $\delta=0$, the resonant result is exactly recovered.
For $\delta\neq0$ we have instead two different regimes, separated by the time $\widetilde t=\Omega/\sigma_\delta^2$ at which $\zeta=1$.
The first one is for $\zeta\ll1$, i.e.~$t\ll\widetilde t$, which produces
\begin{equation}
\begin{split}
    C(t) &\xrightarrow{\zeta\ll1}
    \exp\left[-\frac{\delta^2\sigma_\delta^2t^2}{2\Omega^2}\right] 
    = \exp\left[-\frac{1}{2}\left(\frac{\delta\sigma_\delta t}{\Omega}\right)^2\right] 
    \, , \\
    \varphi(t) &\xrightarrow{\zeta\ll1}
    \frac{(\delta^2+\sigma_\delta^2)t}{2\Omega}\, .
\end{split}
\end{equation}
For $\sigma_\delta\ll|\delta|$ these coincide with the $\lvert\delta\rvert\ll\Omega$ limit of the detuned case in Eq.~\eqref{eq:qs-pr-detuned}, since $\Omega+\delta^2/(2\Omega)\simeq\widetilde\Omega$.
The second regime is for $\zeta\gg1$, i.e.~$t\gg\widetilde t$, which produces
\begin{equation}
\begin{split}
    C(t) &\xrightarrow{\zeta\gg1}
    \sqrt{\frac{\Omega}{\sigma_\delta^2t}}
    \exp\left[-\frac{\delta^2}{2\sigma_\delta^2}\right] \, , \\
    \varphi(t) &\xrightarrow{\zeta\gg1} \frac{\pi}{4} \, .
\end{split}
\end{equation}
In this latter case, the dependence on time of the envelope $C(t)$ is purely algebraic, and originates from those realizations with $\delta_{\rm qs}\simeq-\delta$ that are on resonance, weighted by their probability $\exp[-\delta^2/(2\sigma_\delta^2)]$ (exponentially small for $\sigma_\delta\ll\lvert\delta\rvert$).

We highlight that which of the two regimes actually shapes the damping of the oscillations is determined by the ratio $\sigma_\delta/\lvert\delta\rvert$.
The Gaussian factor can be rewritten as $\exp[-(t/T_{\rm G})^2]$ by introducing $T_{\rm G} = \sqrt2\,\Omega/(\lvert\delta\rvert\sigma_\delta)$, and it can damp the oscillations only for $t\lesssim\widetilde t$, since for $t\gg\widetilde t$ its exponent saturates to $\delta^2/(2\sigma_\delta^2)$.
Since
\begin{equation}
    \frac{T_{\rm G}}{\widetilde t} = \sqrt{2} \frac{\sigma_\delta}{\lvert\delta\rvert} \, ,
\end{equation}
the limit $T_{\rm G}\ll\widetilde t$ (i.e.~$\zeta\ll1$) corresponds to $\sigma_\delta/\lvert\delta\rvert\ll1$, which is exactly the case where the Gaussian envelope dominates and the algebraic tail is exponentially suppressed.
On the other hand, for $\lvert\delta\rvert\lesssim\sigma_\delta$ we have $T_{\rm G}\gtrsim\widetilde t$, so that the Gaussian factor never drops below the constant value $\exp[-\delta^2/(2\sigma_\delta^2)]$: in this case, no Gaussian decay is visible, and the contrast is lost through the algebraic envelope of the $\zeta\gg1$ regime.

\section{Polynomial chaos expansion and Gauss--Hermite quadrature}\label{app:quadrature}
Throughout our optimization routine, we made use of the so-called polynomial chaos expansion (PCE) for the estimation of the moments of the distribution of observables depending on stochastic quantities~\cite{Wiener1938}. Here we provide a brief overview of the method and the protocols we implemented.

Let us consider a model described by a deterministic map $f$ depending on a set of stochastic parameters $\vec\xi=(\xi_1,\dots,\xi_d)$ and providing an output $Y$ (let it be scalar), in formula
\begin{equation}
    Y=f(\vec\xi) \, .
\end{equation}
Our goal is to obtain an estimation of the moments of the distribution of $Y$, which inherits the stochasticity of $\vec\xi$ trough $f$, e.g. we are interested in $\mathbb E[Y]$ or ${\rm var}[Y]$.

Usually, one would run a Monte Carlo simulation by sampling many values $\vec\xi_s$, evolving them through $f$ to obtain many outcomes $Y_s$, and use such outcomes to approximate the distribution of $Y$. This, however, is highly inefficient, especially if the computational cost of evaluating $f$ (which, e.g., in our case implies integrating a Lindblad master equation) is high.

The PCE idea is then to rewrite such quantity by expanding it in a suitable basis of polynomials
\begin{equation}
    Y=f(\vec\xi) = \sum_{\vec\alpha\in{\cal A}}c_{\vec\alpha}\Psi_{\vec\alpha}(\vec\xi)
    \label{eq:PCE}
\end{equation}
which, in case of independent random variables $\xi_j$, further simplifies into
\begin{equation}
    Y = \sum_{\vec\alpha\in{\cal A}} \prod_{j=1}^d c_{\alpha_j} \psi ^{(j)}_{\alpha_j} (\xi_j) \, ,
\end{equation}
where $c_{\vec\alpha}$ are deterministic coefficients given by
\begin{equation}
    c_{\vec\alpha} = \frac{\mathbb E [f(\vec\xi) \Psi_{\vec\alpha}(\vec\xi)]}{\mathbb E[\Psi_{\vec\alpha}^2(\vec\xi)]}\, ,
    \label{eq:PCE-coeffs}
\end{equation}
$\Psi_{\vec\alpha}$ are multivariate polynomials chosen according to the distribution of $\vec\xi$, and ${\cal A}$ is the set enumerating elements of the chosen basis of polynomials. For instance, in case of a random variable $\xi$ distributed according to a Gaussian, a common choice is to use Hermite polynomials, which satisfy orthogonality condition with respect to such distribution.
Moments of the distribution of $Y$ can thus be computed in terms of the $c_{\vec\alpha}$ by noting that $Y=c_0 + \sum_{\vec\alpha\neq0}c_{\vec\alpha}\Psi_{\vec\alpha}$, in particular we have (for normalized polynomials)
\begin{align}
    \mathbb E[Y] &= c_0 \, , \\
    \mathbb {\rm var}[Y] &= \sum_{\vec\alpha\neq0}c_{\vec\alpha}^2 \, .
\end{align}

Since storing an infinite number of terms in the expansion of Eq.~\eqref{eq:PCE} is numerically impractical, one usually truncates the polynomial basis to a maximum order $P$, thus providing an approximation $Y_P$ of $Y$.

One might wonder at this point how to numerically obtain an estimation of the expected values in Eq.~\eqref{eq:PCE-coeffs} necessary for computing the coefficients entering the PCE.
In the case of Gaussian distribution of the stochastic variables, one can use the Gauss--Hermite quadrature~\cite{Golub1969}.
Let us consider for simplicity the case of a single stochastic input sampled from a normal distribution $\xi\sim{\cal N}(0,1)$. The expected value of $f(\xi)$ is
\begin{equation}
    \mathbb E[f(\xi)]=\int_{-\infty}^\infty \frac{1}{\sqrt{2\pi}} e^{-x^2/2} f(x) {\rm d}x \, .
\end{equation}
The Gauss--Hermite quadrature allows us to approximate such integral by computing values of $f$ in a finite amount $Q$ of points, so-called quadrature nodes, and weighting them appropriately, namely
\begin{equation}
    \mathbb E[f(\xi)] \approx \sum_{q=0}^Q w_q f(x_q)
    \label{eq:gauss-hermite-mean}
\end{equation}
where $x_q$ are the roots of the of probabilist's Hermite polynomials ${\rm He}_Q(x)$ defined by the recurrence relation
\begin{equation}
\begin{split}
    &{\rm He}_{q+1}(x) = x{\rm He}_q(x) - q{\rm He}_{q-1}(x)\\
    &\text{with }{\rm He}_0(x)=1 \text{ and } {\rm He}_1(x)=x,
\end{split}
\end{equation}
and the corresponding weights are
\begin{equation}
    w_q=\frac{Q!}{Q^2{\rm He}_{q-1}^2(x_q)} \, .
\end{equation}
We recall that $Q+1$ quadrature nodes are sufficient to have a precise estimation of a polynomial of degree $Q$.
The PCE coefficients in Eq.~\eqref{eq:PCE-coeffs} are then approximated as
\begin{equation}
    c_\alpha \approx \sum_{q=0}^Q w_q f(x_q) \psi_\alpha(x_q) \, .
    \label{eq:PCE-coeffs-quad}
\end{equation}

The protocol we employed in our work is then the following. We describe it for the relevant example of quasi-static detuning noise for a single qubit, but can be easily adapted to other cases.
The stochastic variable under consideration is $\delta_{\rm qs}\sim{\cal N}(\delta_0,\sigma_\delta^2)=\delta_0+\sigma_\delta\xi$, with $\xi\sim{\cal N}(0,1)$, while the outcome is the probability of Rydberg-level population $o=\tr[n_z\rho]$. Steps are the following:
\begin{enumerate}
    \item choose a maximum PCE degree $P$ and number of quadrature nodes $Q\geq P+1$;
    \item get Gauss--Hermite quadrature nodes and weights $\xi_q, w_q$;
    \item integrate the Lindblad master equation for each quasi-static detuning noise value $\delta_q=\delta_0 + \sigma_\delta\xi_q$ to get density matrix $\rho_q$ and obtain the respective observable value $o_q=\tr[n_z \rho_q]$;
    \item compute PCE coefficients $c_p$ using Eq.~\eqref{eq:PCE-coeffs-quad};
    \item estimate $\widehat o=c_0$, $\widehat{\rm var}[o]=\sum_{p=1}^Pc_p^2$.
\end{enumerate}

In our experience, we have seen good agreement of the PCE estimates with Monte Carlo ones already using few tens of quadrature nodes, in contrast with thousands of evaluations required by the latter.

We note that PCE is not needed when the model contains a single stochastic variable, since Gauss--Hermite quadrature already suffices: the mean $\mathbb{E}[f(\xi)]$ is given by Eq.~\eqref{eq:gauss-hermite-mean}, and the variance by
\begin{equation}
    {\rm var}[f(\xi)] = \sum_q w_q f^2(\xi_q) - \left( \sum_q w_q f(\xi_q) \right)^2 \, .
\end{equation}
PCE becomes advantageous when the model contains several stochastic variables.
For $Q$ quadrature points per variable and $S$ stochastic variables, a tensor-product Gauss--Hermite quadrature requires $Q^S$ evaluations, which grows exponentially with $S$. In contrast, a PCE truncated at total order $P$ involves only $\binom{S+P}{P}$ coefficients, which is typically more manageable.

\section{Bayesian estimation of statistical errors}\label{app:bayes}
We present here the procedure used to estimate the statistical uncertainties associated with the quantities reported in this paper.
The protocol relies on Bayesian inference to determine the posterior distribution of a given parameter from observed data~\cite{Gelman2014,Hall2021}.

In particular, let us assume that we observe a binary random variable $\xi\in\{0,1\}$ (e.g. the the measurement outcome of a single-qubit Pauli operator, appropriately mapped) and wish to estimate its associated probability of success $p_0$ from repeated independent samples.
The likelihood of observing $N_0$ successes in $N_s$ trials, given $p_0$, is described by the binomial distribution:
\begin{equation}
    \mathbb{P}(N_0 \mid N_s,p_0) = \binom{N_s}{N_0} p_0^{N_0} (1-p_0)^{N_s-N_0} \, .
\end{equation}
Bayes' theorem then gives the posterior distribution of $p_0$ conditioned on the observed data:
\begin{equation}
\begin{split}
    \mathbb{P}(p_0 \mid N_0,N_s) 
    &= \frac{ \mathbb{P}(N_0 \mid N_s,p_0)\, \mathbb{P}(p_0) }{ \mathbb{P}(N_0 \mid N_s) } \\
    &\propto \mathbb{P}(N_0 \mid N_s,p_0)\, \mathbb{P}(p_0) \, ,
\end{split}
\end{equation}
where the evidence $\mathbb{P}(N_0 \mid N_s)$ ensures the proper normalization of the posterior distribution, and $\mathbb{P}(p_0)$ represents our prior knowledge about $p_0$ before observing the data.

For a binomial likelihood, the conjugate prior is the Beta distribution,
\begin{equation}
\begin{split}
    \mathbb{P}(p_0)=f_{\mathrm{B}}(p_0 \mid \alpha,\beta) &= \frac{ p_0^{\alpha-1}(1-p_0)^{\beta-1} }{ \mathrm{B}(\alpha,\beta) } \\ &= \frac{ \Gamma(\alpha+\beta) }{ \Gamma(\alpha)\Gamma(\beta) } p_0^{\alpha-1}(1-p_0)^{\beta-1} \, ,
\end{split}
\end{equation}
where
\begin{equation}
\mathrm{B}(\alpha,\beta) = \frac{ \Gamma(\alpha)\Gamma(\beta) }{ \Gamma(\alpha+\beta) }
\end{equation}
is the Beta function, and $\Gamma$ denotes the Gamma function.
Since, \textit{a priori}, we assume no specific knowledge about $p_0$, a natural choice is a uniform prior over the interval $[0,1]$, which corresponds to setting $\alpha=\beta=1$.

The advantage of using a conjugate prior, is that the posterior will belong to the same family of distributions. Hence, since the Beta prior is conjugate to the binomial likelihood, the posterior distribution also belongs to the Beta family:
\begin{equation}
\begin{split}
\mathbb{P}(p_0 \mid N_0,N_s) 
&= \binom{N_s}{N_0} \frac{p_0^{N_0+\alpha-1} (1-p_0)^{N_s-N_0+\beta-1}}{ \mathrm{B}(\alpha,\beta) }\\
&\vphantom{\frac11}\propto f_{\mathrm{B}} \left( p_0 \mid \alpha+N_0,\, \beta+N_s-N_0 \right)
\, .
\end{split}
\end{equation}

The posterior mean can thus be used as a Bayesian point estimator of $p_0$:
\begin{equation}
\widehat{p}_0 = \frac{N_0+\alpha}{N_s+\alpha+\beta} \xrightarrow{\alpha=\beta=1} \frac{N_0+1}{N_s+2} \, ,
\end{equation}
which may be interpreted as incorporating $\alpha$ prior successes and $\beta$ prior failures into the observed data.
A credible interval for $p_0$ can be constructed accordingly from the quantiles of the posterior distribution; for example, an equal-tailed $95\%$ credible interval is obtained from the $2.5\%$ and $97.5\%$ quantiles.

From the posterior distribution, credible intervals for quantities defined as (potentially nonlinear) functions $g(p_0)$ can be obtained as well. The simplest approach is to draw many samples $p_0^{(r)} \sim \mathbb{P}(p_0 \mid N_0,N_s)$ and compute the corresponding values $g(p_0^{(r)})$. This yields an empirical approximation of the distribution of $g(p_0)$, from which its quantiles, and hence credible intervals, can be estimated in a Monte Carlo fashion.

In the case of random variables $\xi$ distributed according to a multinomial, instead (e.g. measurement of multiqubit operators), the same approach can be straightforwardly generalized. Let $\xi\in\{1,\dots,K\}$, the likelihood of observing a collection of samples $N_c=(N_1,\dots,N_K)$, distributed over the classes with probabilities $p_c=(p_1,\dots,p_K)$, out of $N_s$ total ones, is
\begin{equation}
    \mathbb{P}(N_c\mid N_s,p_c) = {N_s!} \prod_{k=1}^K \frac{p_k^{N_k}}{N_k!} \, ,
\end{equation}
with the conditions $\sum_{k=1}^K N_k=N_s$ and $\sum_{k=1}^K p_k = 1$.
The conjugate prior is in this case a Dirichlet distribution
\begin{equation}
    \mathbb{P}(p_c)=f_{\rm D}(p_c\mid \alpha_c) = \Gamma(\alpha_0) \prod_{k=1}^K \frac{p_k^{\alpha_k-1}}{\Gamma(\alpha_k)} \, ,
\end{equation}
with $\alpha_c=(\alpha_1,\dots,\alpha_K)$ and $\alpha_0=\sum_{k=1}^K\alpha_k$, which equals the uniform distribution for $\alpha_k=1 \,\forall\,k$.
The posterior distribution then follows as
\begin{equation}
\begin{split}
    \mathbb{P}(p_c\mid N_c,N_s) 
    &= N_s!\:\Gamma(\alpha_0)\prod_{k=1}^K \frac{p_k^{N_k+\alpha_k-1}}{N_k!\:\Gamma(\alpha_k)}\\
    &\propto\vphantom{\frac11}f_{\rm D}(p_c\mid \alpha_c + N_c) \, ,
\end{split}
\end{equation}
where we denoted $\alpha_c + N_c=(\alpha_1 + N_1,\dots, \alpha_K + N_K)$,
the posterior mean is
\begin{equation}
    \widehat p_k = \frac{N_k+\alpha_k}{N_s+\alpha_0} \xrightarrow{\alpha_k=1\,\forall\,k} \frac{N_k+1}{N_s+K} \, ,
\end{equation}
and its marginal distribution is
\begin{equation}
\begin{split}
    f_{\rm B}(p_k\mid&\alpha_k,\alpha_0-\alpha_k)\\
    &\xrightarrow{\alpha_k=1\,\forall\,k}
    f_{\rm B}(p_k\mid N_k+1,N_s+K-N_k-1) \, ,
\end{split}
\end{equation}
namely the distribution describing the binary case of being inside or outside of the $k$-th class. This latter, in particular, is the distribution used for constructing the credible interval associated to each $p_k$.
It is easy to see that the binomial case is recovered for $K=2$.

Let us now exemplify, for better clarity, this approach in the case of computing the statistical errors associated to the computational states population averaged over (cluster of) atoms' positions.
Since we perform this operation both for one- and two-qubit populations, we will address directly the more general multinomial case.
In this case, $N_s$ denotes the number of shots performed, and $N_k$ the number of times the $k$-th bitstring has been measured.
Since we assume the measurement process independent for each cluster of atoms, the likelihood factorizes in a product of multinomials
\begin{equation}
    \mathbb{P}(\vec{N_c}\mid {N_s},\vec{p_c}) = \prod_{j=1}^M\left({N_s!} \prod_{k=1}^K \frac{p_{k,j}^{N_{k,j}}}{N_{k,j}!}\right) \, ,
\end{equation}
and so do the prior and posterior distributions as well
\begin{align}
    \mathbb{P}(\vec{p_c}) &= \prod_{j=1}^M f_{\mathrm{D}}(p_{c,j} \mid \alpha_{c,j}) \, ,\\
    \mathbb{P}(\vec{p_c} \mid \vec{N_c},{N_s}) &= \prod_{j=1}^M f_{\mathrm{D}}(p_{c,j} \mid \alpha_{c,j}+N_{c,j}) \, ,
\end{align}
where the vectorial notation describes the collection of quantities realated to each cluster of atoms, the subscript $j$ denotes the quantities related to the $j$-th cluster, and $M$ the number of clusters (i.e.~$M=N$ for single-qubit experiments and $M=N/2$ for two-qubit experiments).
We are now interested in computing the average computational states population $\widehat p_c^{\rm avg}$ over atoms' positions. Owing to the linearity of the average and the independence of samples, we can write
\begin{equation}
    \widehat p_c^{\rm avg} = \frac{1}{M}\sum_{j=1}^M \widehat p_{c,j} = \frac{1}{M}\sum_{j=1}^M \frac{N_{c,j}+\alpha_{c,j}}{N_{s}+\alpha_{0,j}} \, .
\end{equation}
The empirical distribution of the average $\mathbb{P}(p_c^{\rm avg} \mid \vec{N_c},{N_s})$ and its quantiles, thus the credible interval associated to $\widehat p_c^{\rm avg}$, can be reconstructed by means of samples
\begin{equation}
    \widehat p_c^{{\rm avg},(r)} = \frac{1}{M}\sum_{j=1}^M \widehat p_{c,j}^{(r)} \, ,
\end{equation}
where $\widehat p_{c,j}^{(r)}\sim \mathbb{P}(p_{c,j} \mid N_{c,j},N_{s}) \propto f_{\rm D}(p_{c,j}\mid \alpha_{c,j} + N_{c,k})$.

\section{Extended experimental data and results}\label{app:extended-data}

We provide in this Appendix supplementary plots for experiments not displayed in the main text.
Relevant comments about them are, in this case, relegated to plots' captions.

\begin{figure*}
    \centering
    \begin{minipage}[c]{0.48\textwidth}
        \centering
        \includegraphics[width=0.95\linewidth]{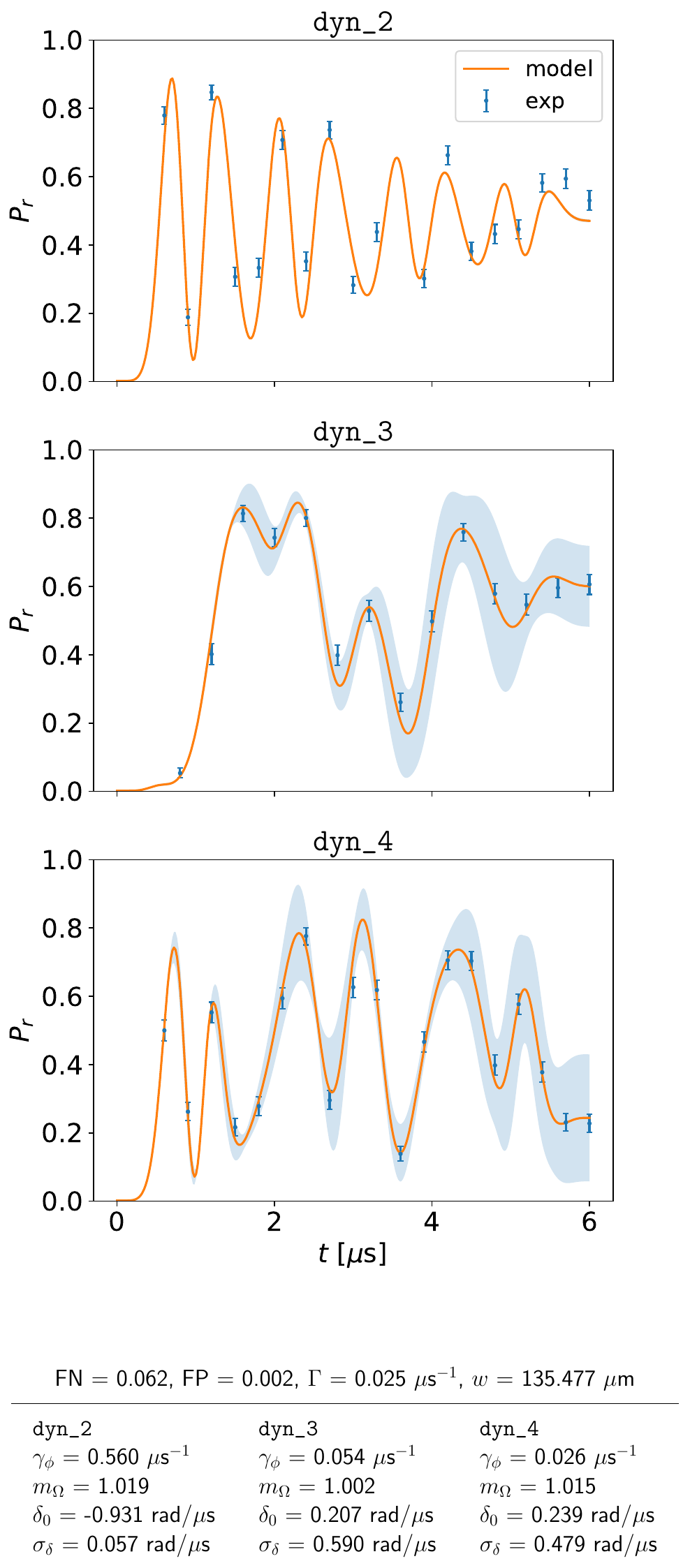}
        \cprotect\caption{Comparison between the experimental data obtained from datasets \verb|dyn_2|, \verb|dyb_3|, and and \verb|dyn_4| (averaged over atoms' positions) and the predictions of our model with optimized parameters. Parameters listed above the horizontal line are shared between datasets, while those listed below it are fitted separately for each dataset, and yield $\oRMSZ\approx0.811$ and $\oRMSE\approx0.098$. We note that dataset \verb|dyn_2| is reproduced less accurately than the other two, as reflected in the overall higher $\oRMSZ$ and $\oRMSE$. We attribute this to an interruption during its acquisition, when the hardware was recalibrated. We therefore fitted our model to \verb|dyn_3| and \verb|dyn_4| only, obtaining better results (cf. Fig.~\ref{fig:dyn_3-4}). Error bars on the experimental data are obtained with the Bayesian estimation procedure described in App.~\ref{app:bayes}. The shaded area indicates the one-standard-deviation spread of the model prediction arising from the quasi-static detuning distribution with width $\sigma_\delta$.}
        \label{fig:dyn_2-3-4}
    \end{minipage}\hfill
    \begin{minipage}[c]{0.48\textwidth}
        \centering
        \includegraphics[width=0.95\linewidth]{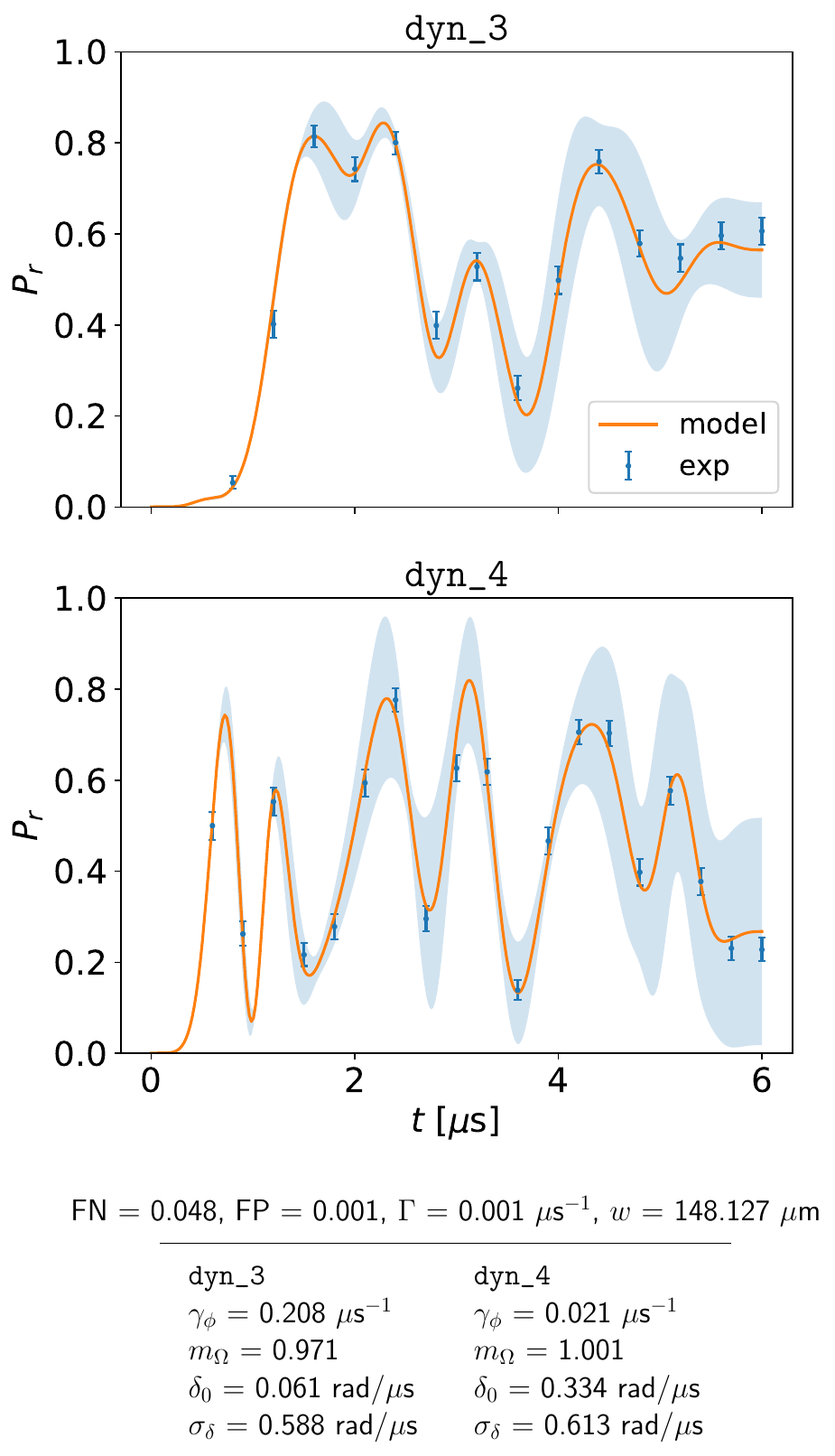}
        \cprotect\caption{Comparison between the experimental data obtained from datasets \verb|dyn_3| and \verb|dyn_4| (averaged over atoms' positions) and the predictions of our model with optimized parameters. Parameters listed above the horizontal line are shared between datasets, while those listed below it are fitted separately for each dataset, and yield $\oRMSZ\approx0.637$ and $\oRMSE\approx0.079$. Error bars on the experimental data are obtained with the Bayesian estimation procedure described in App.~\ref{app:bayes}. The shaded area indicates the one-standard-deviation spread of the model prediction arising from the quasi-static detuning distribution with width $\sigma_\delta$.}
        \label{fig:dyn_3-4}
    \end{minipage}
\end{figure*}

\begin{figure*}
    \centering
    \includegraphics[width=0.95\linewidth]{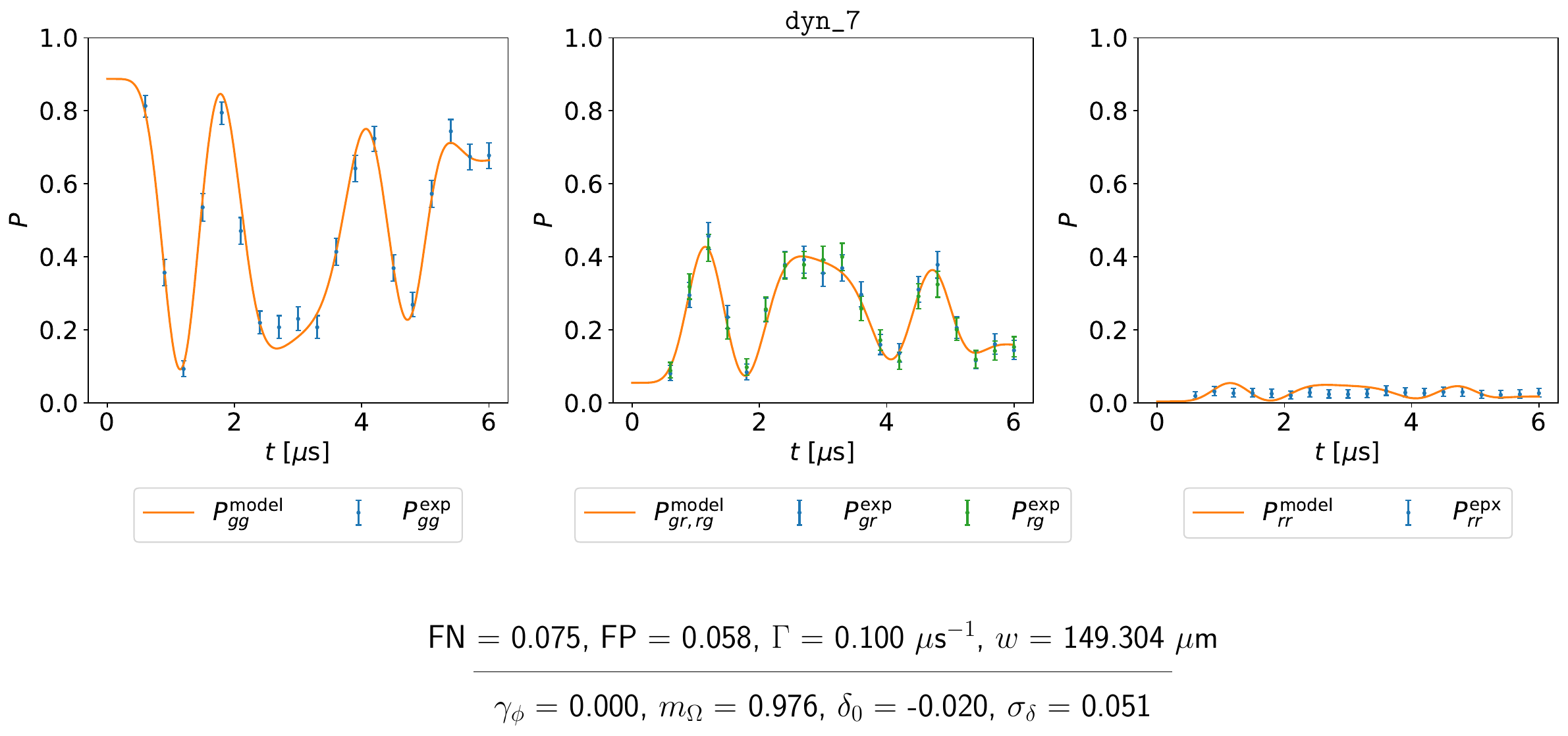}
    \cprotect\caption{Comparison between the experimental data obtained from dataset \verb|dyn_7| (averaged over pairs of atoms) and the predictions of our model with optimized parameters, yielding $\oRMSZ\approx0.496$ and $\oRMSE\approx0.048$. Error bars on the experimental data are obtained with the Bayesian estimation procedure described in App.~\ref{app:bayes}. The shaded area indicates the one-standard-deviation spread of the model prediction arising from the quasi-static detuning distribution with width $\sigma_\delta$.}
    \label{fig:dyn_7}
\end{figure*}

\begin{figure*}
    \centering
    \includegraphics[width=0.95\linewidth]{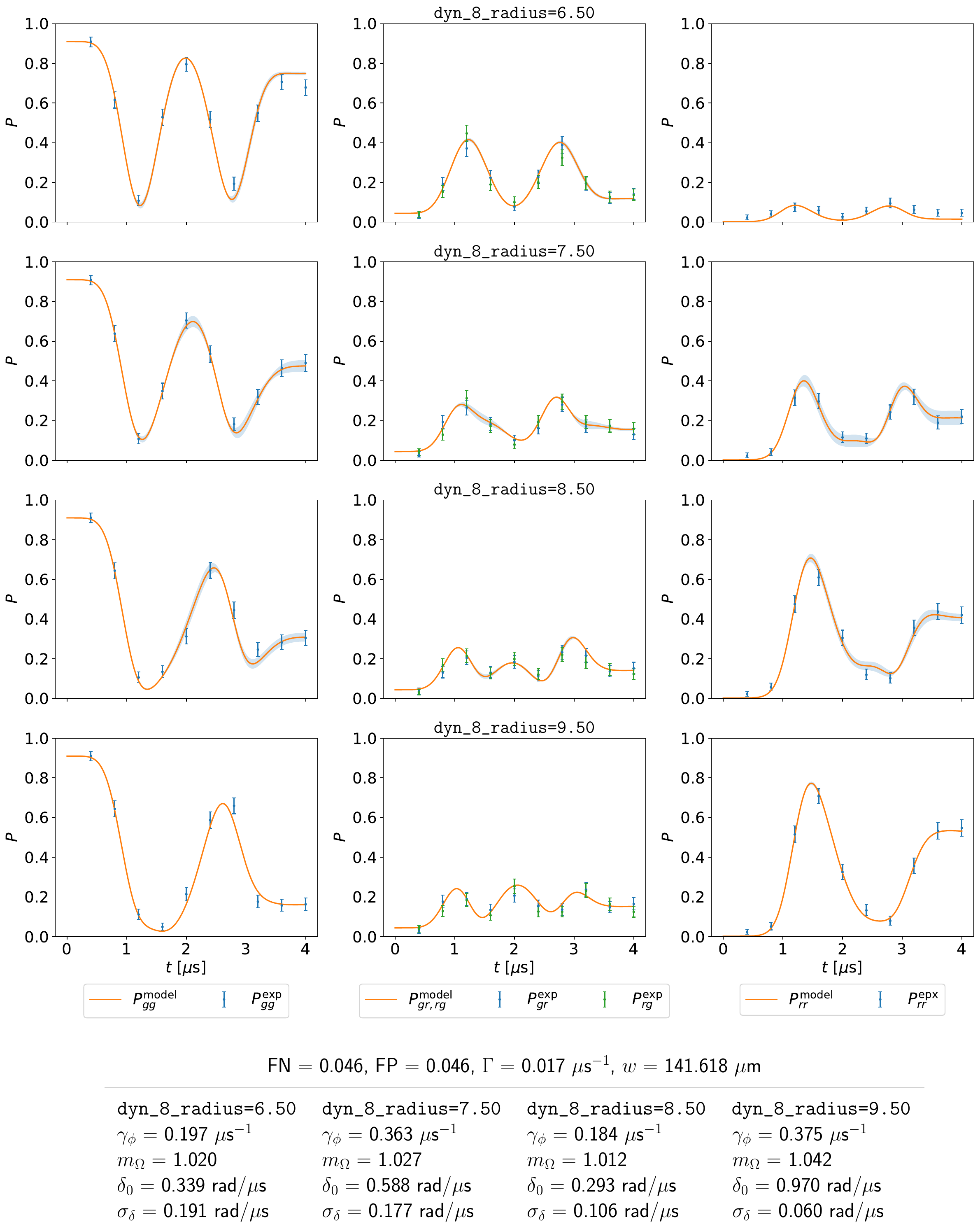}
    \cprotect\caption{Comparison between the experimental data obtained from datasets \verb|dyn_8| with various distances between atoms (averaged over pairs of atoms) and the predictions of our model with optimized parameters. Parameters listed above the horizontal line are shared between datasets, while those listed below it are fitted separately for each dataset, and yield $\oRMSZ\approx0.520$ and $\oRMSE\approx0.054$. Error bars on the experimental data are obtained with the Bayesian estimation procedure described in App.~\ref{app:bayes}. The shaded area indicates the one-standard-deviation spread of the model prediction arising from the quasi-static detuning distribution with width $\sigma_\delta$.}
    \label{fig:dyn_8}
\end{figure*}

\begin{figure*}
    \centering
    \includegraphics[width=0.95\linewidth]{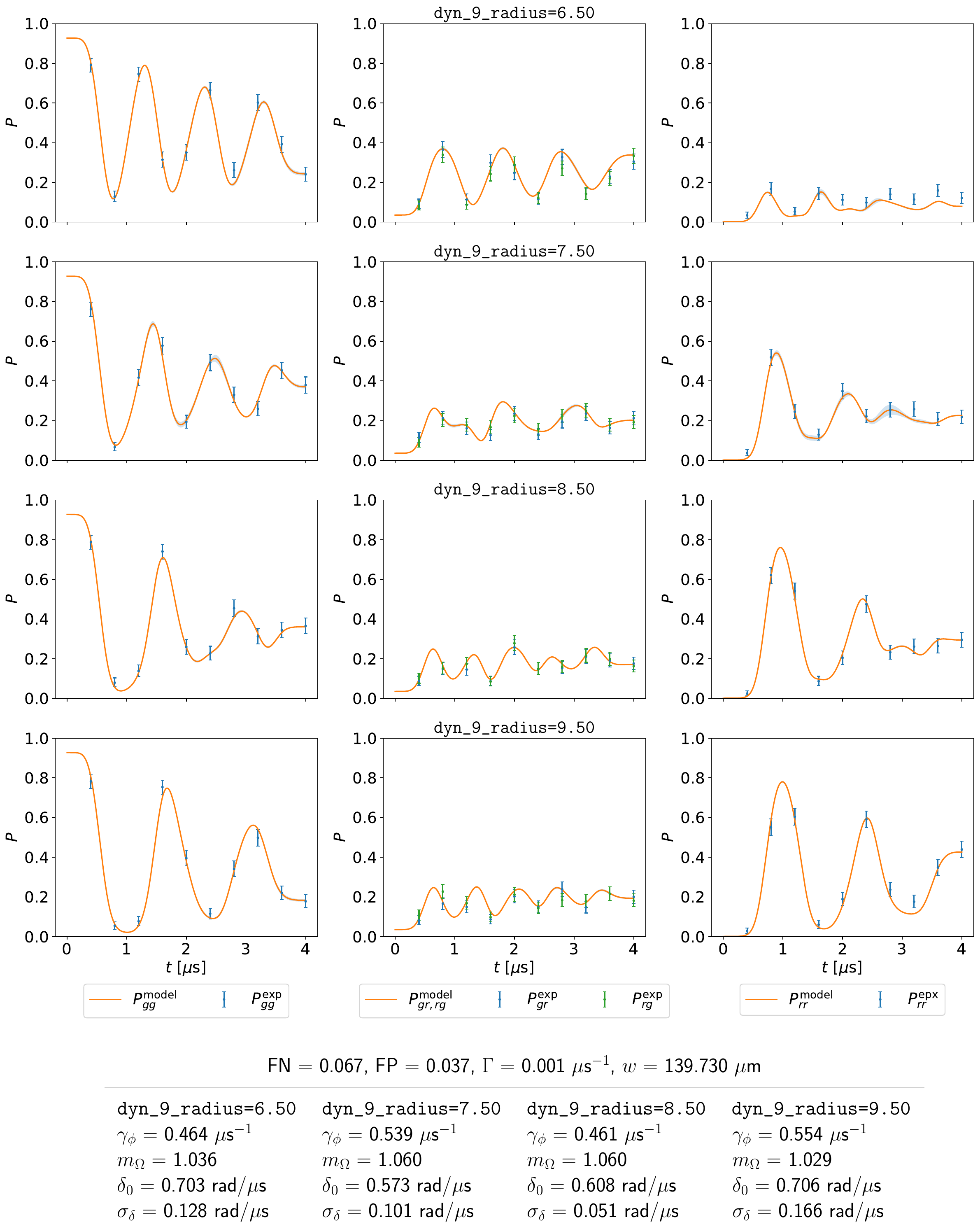}
    \cprotect\caption{Comparison between the experimental data obtained from datasets \verb|dyn_9| with various distances between atoms (averaged over pairs of atoms) and the predictions of our model with optimized parameters. Parameters listed above the horizontal line are shared between datasets, while those listed below it are fitted separately for each dataset, and yield $\oRMSZ\approx0.529$ and $\oRMSE\approx0.059$. Error bars on the experimental data are obtained with the Bayesian estimation procedure described in App.~\ref{app:bayes}. The shaded area indicates the one-standard-deviation spread of the model prediction arising from the quasi-static detuning distribution with width $\sigma_\delta$.}
    \label{fig:dyn_9}
\end{figure*}

\begin{figure*}
    \centering
    \includegraphics[width=0.95\linewidth]{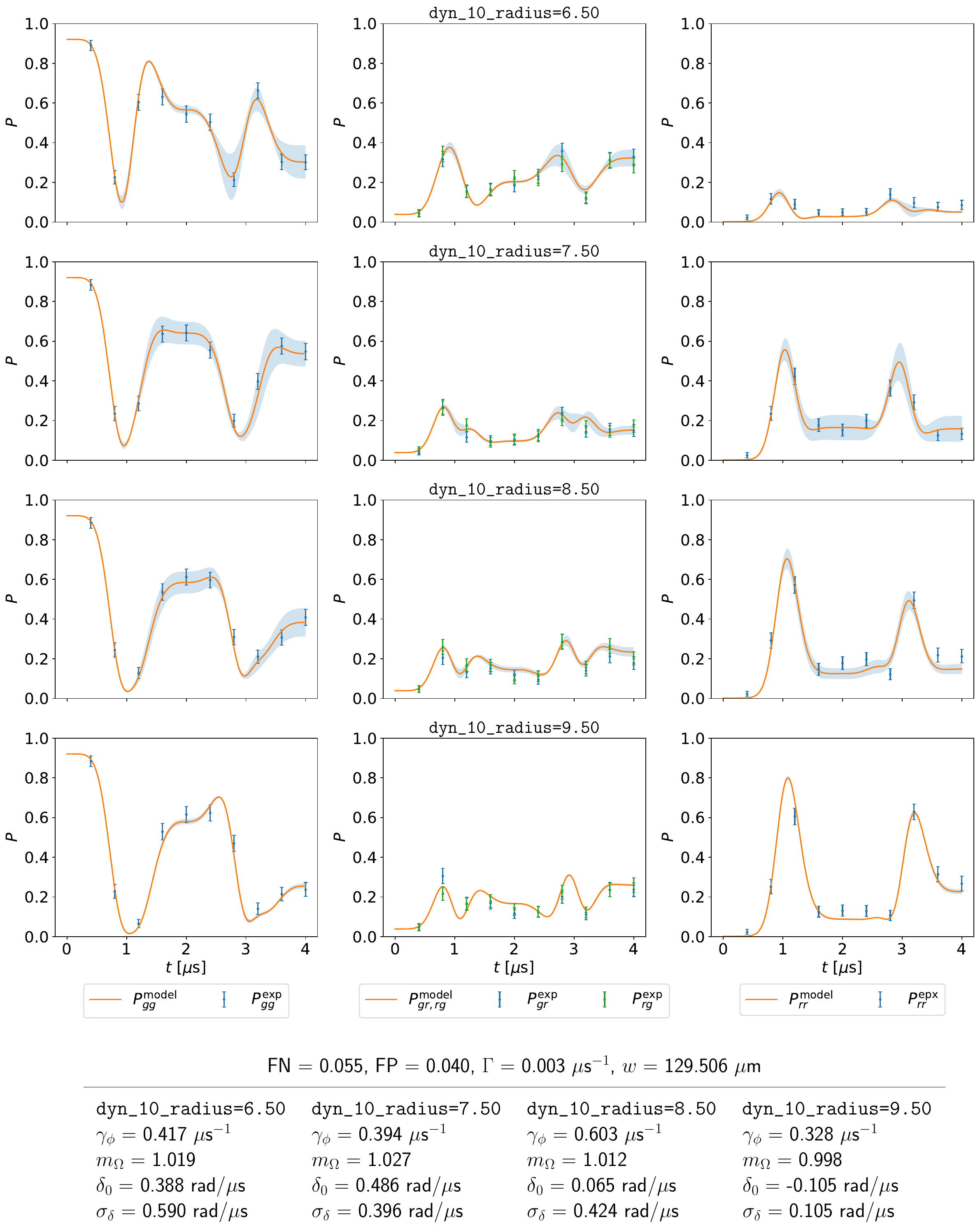}
    \cprotect\caption{Comparison between the experimental data obtained from datasets \verb|dyn_10| with various distances between atoms (averaged over pairs of atoms) and the predictions of our model with optimized parameters. Parameters listed above the horizontal line are shared between datasets, while those listed below it are fitted separately for each dataset, and yield $\oRMSZ\approx0.560$ and $\oRMSE\approx0.061$. Error bars on the experimental data are obtained with the Bayesian estimation procedure described in App.~\ref{app:bayes}. The shaded area indicates the one-standard-deviation spread of the model prediction arising from the quasi-static detuning distribution with width $\sigma_\delta$.}
    \label{fig:dyn_10}
\end{figure*}

\begin{figure*}
    \centering
    \includegraphics[width=0.95\linewidth]{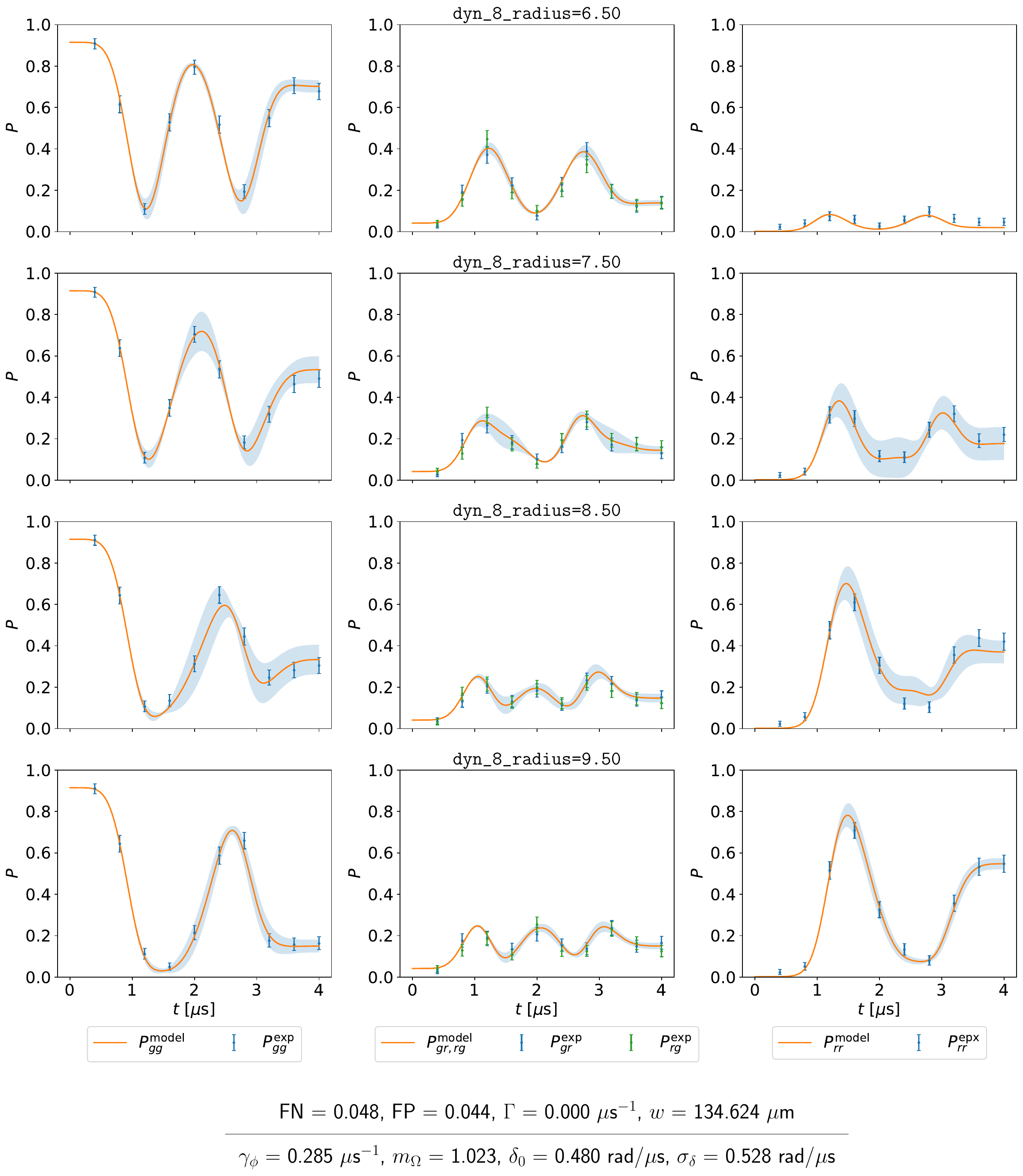}
    \cprotect\caption{Comparison between the experimental data obtained from datasets \verb|dyn_8| with various distances between atoms (averaged over pairs of atoms) and the predictions of our model with optimized parameters. All parameters are shared between datasets, and yield $\oRMSZ\approx0.506$ and $\oRMSE\approx0.053$. Error bars on the experimental data are obtained with the Bayesian estimation procedure described in App.~\ref{app:bayes}. The shaded area indicates the one-standard-deviation spread of the model prediction arising from the quasi-static detuning distribution with width $\sigma_\delta$.}
    \label{fig:dyn_8-common}
\end{figure*}

\begin{figure*}
    \centering
    \includegraphics[width=0.95\linewidth]{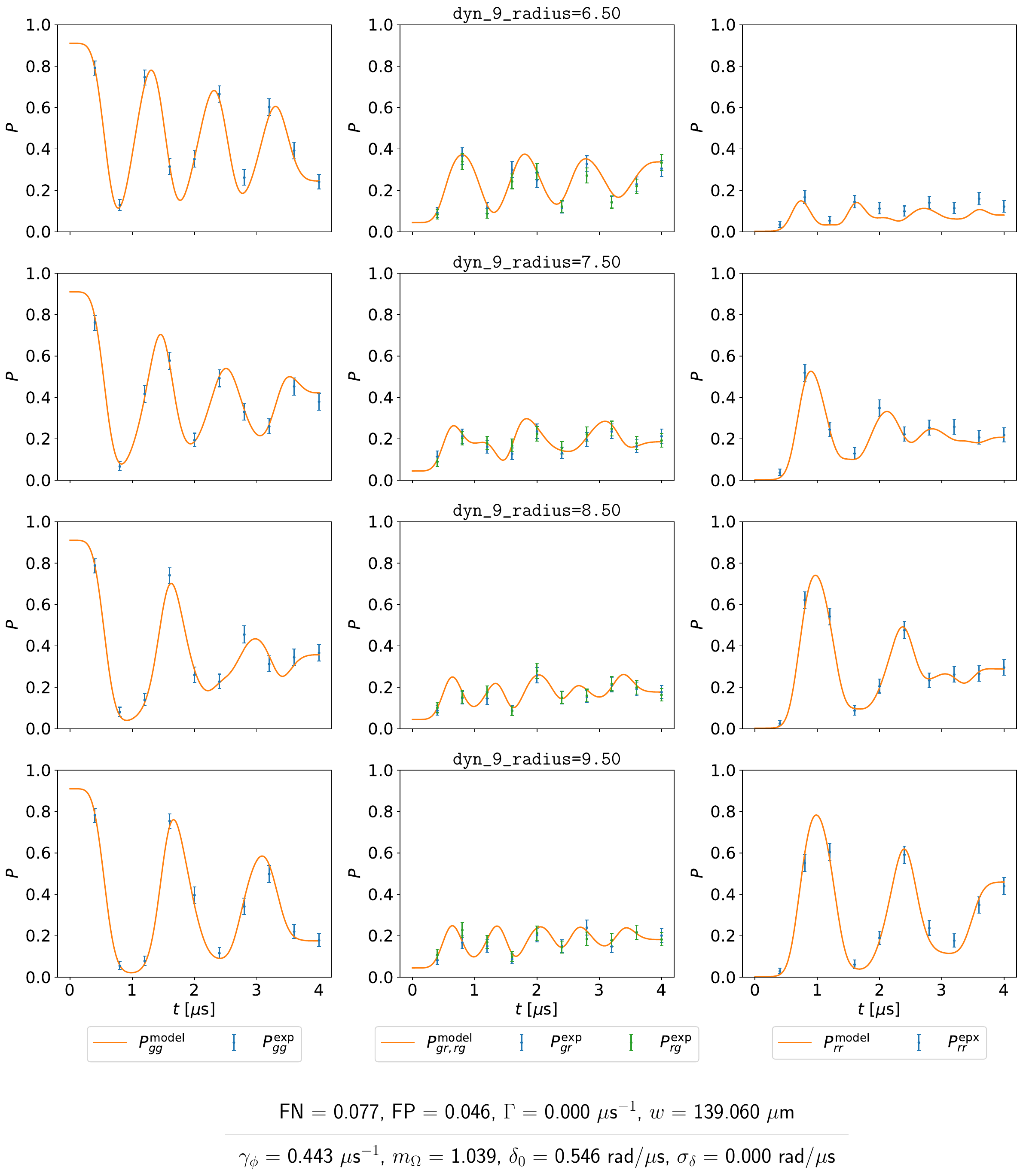}
    \cprotect\caption{Comparison between the experimental data obtained from datasets \verb|dyn_9| with various distances between atoms (averaged over pairs of atoms) and the predictions of our model with optimized parameters. All parameters are shared between datasets, and yield $\oRMSZ\approx0.533$ and $\oRMSE\approx0.060$. Error bars on the experimental data are obtained with the Bayesian estimation procedure described in App.~\ref{app:bayes}. The shaded area indicates the one-standard-deviation spread of the model prediction arising from the quasi-static detuning distribution with width $\sigma_\delta$.}
    \label{fig:dyn_9-common}
\end{figure*}

\begin{figure*}
    \centering
    \includegraphics[width=0.95\linewidth]{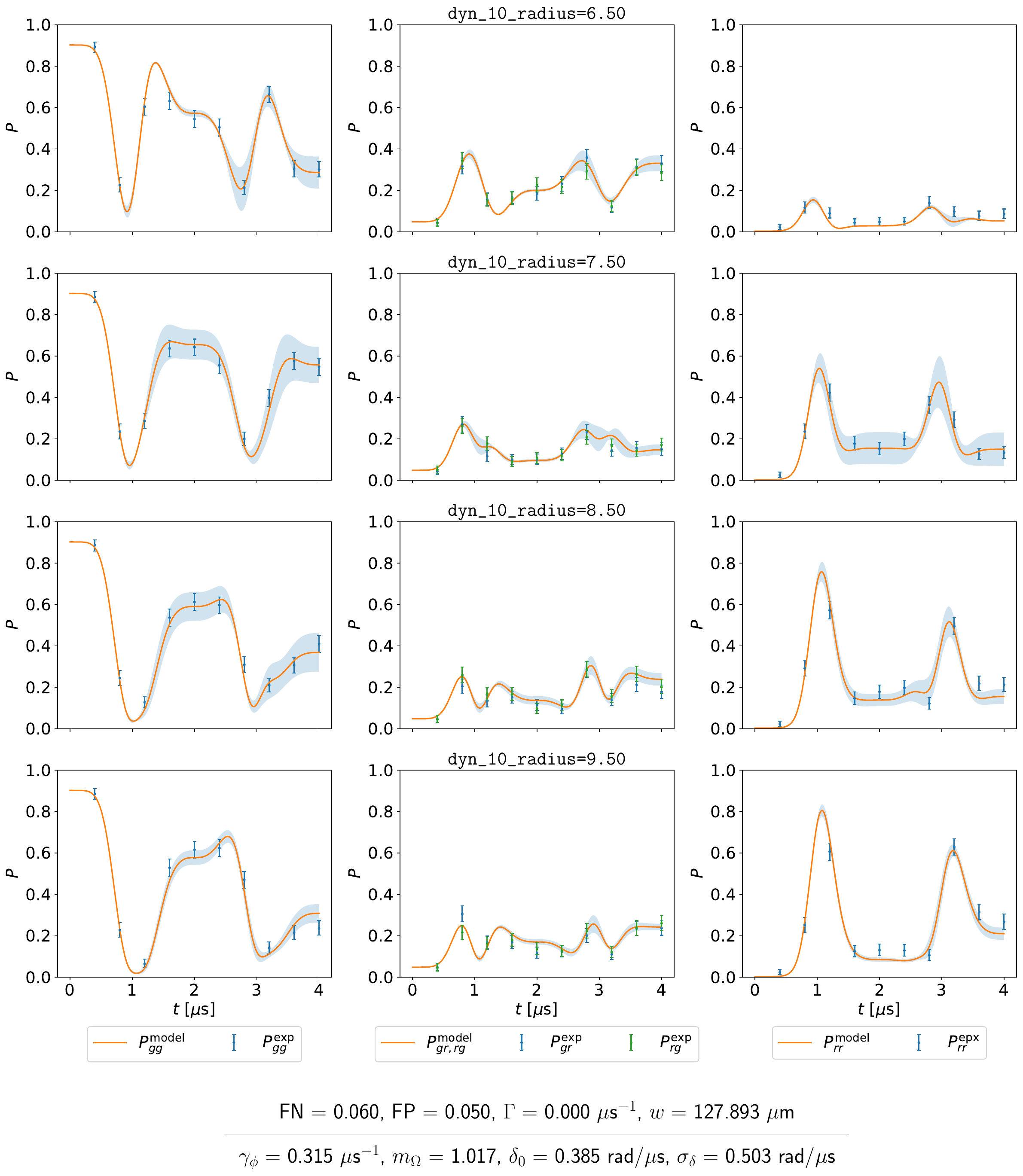}
    \cprotect\caption{Comparison between the experimental data obtained from datasets \verb|dyn_10| with various distances between atoms (averaged over pairs of atoms) and the predictions of our model with optimized parameters. All parameters are shared between datasets, and yield $\oRMSZ\approx0.558$ and $\oRMSE\approx0.061$. Error bars on the experimental data are obtained with the Bayesian estimation procedure described in App.~\ref{app:bayes}. The shaded area indicates the one-standard-deviation spread of the model prediction arising from the quasi-static detuning distribution with width $\sigma_\delta$.}
    \label{fig:dyn_10-common}
\end{figure*}

\begin{figure}
    \centering
    \includegraphics[width=0.95\linewidth]{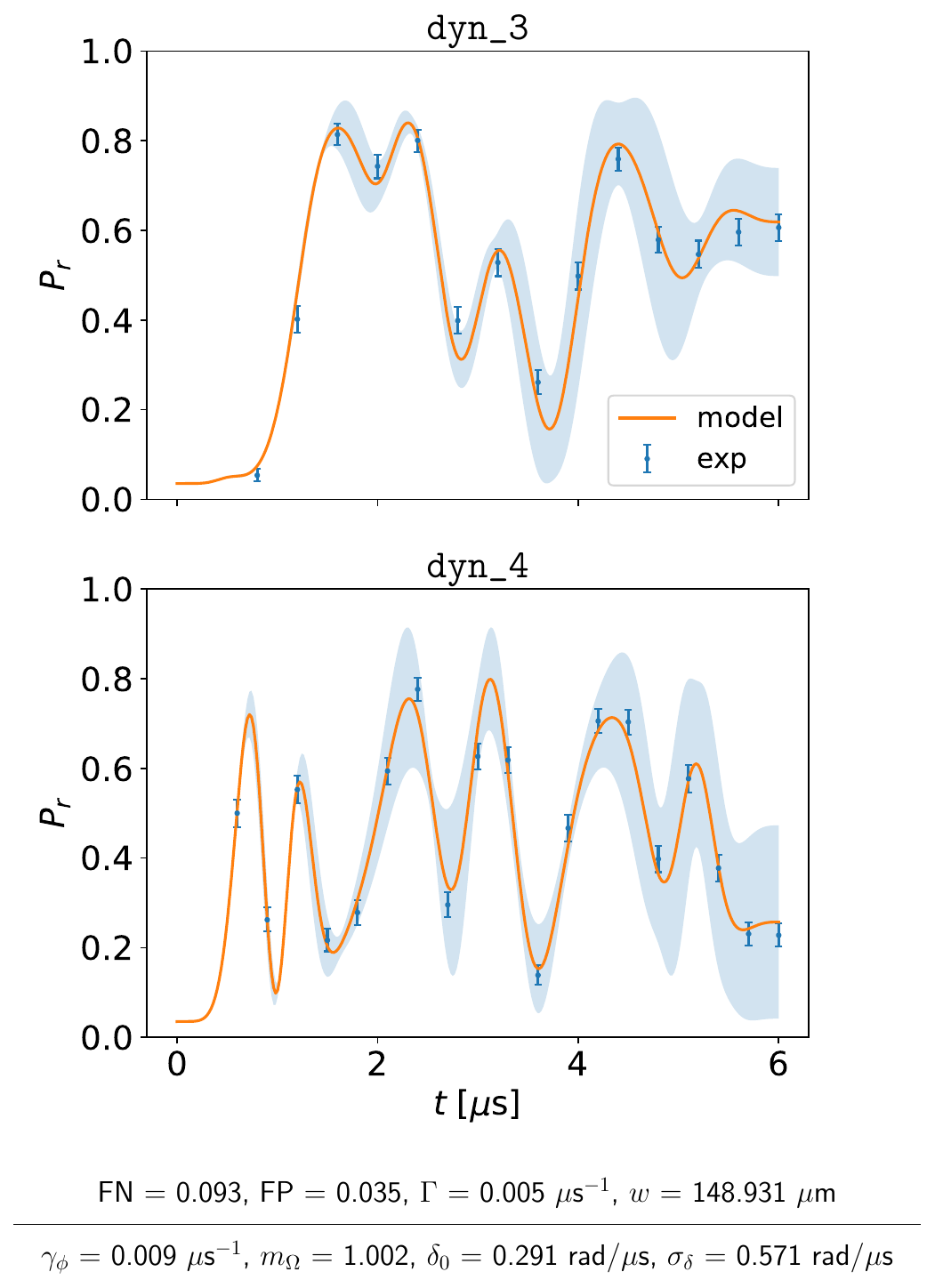}
    \cprotect\caption{Comparison between the experimental data obtained from datasets \verb|dyn_3| and \verb|dyn_4| (averaged over atoms' positions) and the predictions of our model with optimized parameters. All parameters are shared between datasets, and yield $\oRMSZ\approx0.645$ and $\oRMSE\approx0.080$. Error bars on the experimental data are obtained with the Bayesian estimation procedure described in App.~\ref{app:bayes}. The shaded area indicates the one-standard-deviation spread of the model prediction arising from the quasi-static detuning distribution with width $\sigma_\delta$.}
    \label{fig:dyn_3-4-common}
\end{figure}

\bibliography{refs}

\end{document}